\documentclass[a4paper]{article}
\usepackage{graphicx}
\usepackage{bm}
\usepackage{dcolumn}
\usepackage{amsmath}
\usepackage{amssymb}
\begin{document}

\title{Model error and sequential data assimilation. A deterministic formulation}

\author{A. Carrassi$^*$, S. Vannitsem, and C. Nicolis}
\maketitle
\begin{center}
Institut Royal M\'et\'eorologique de Belgique, Bruxelles, Belgique$^*$
\end{center}

{\small $^*$Corresponding author e-mail: a.carrassi@oma.be}

\begin{abstract}

Data assimilation schemes are confronted with the presence of model errors arising from the imperfect description of atmospheric dynamics. 
These errors are usually modeled on the basis of simple assumptions such as bias, white noise, first order Markov process. In the present work, a 
formulation of the sequential extended Kalman filter is proposed, based on recent findings on the universal deterministic behavior of model errors 
in deep contrast with previous approaches (Nicolis, 2004).
This new scheme is applied in the context of a spatially distributed system proposed by Lorenz (1996).
It is found that (i) for short times, the estimation error is accurately approximated by an evolution law in which the variance of the model error
(assumed to be a deterministic process) evolves according to a quadratic law, in agreement with the theory. Moreover, the correlation 
with the initial condition error appears to 
play a secondary role in the short time dynamics of the estimation error covariance. (ii) The deterministic description of the model error evolution, 
incorporated into the classical extended Kalman filter equations, reveals that substantial improvements of the filter accuracy can be gained 
as compared with the classical white noise assumption.
The universal, short time, quadratic law for the evolution of the model error covariance matrix seems very promising for modeling estimation error 
dynamics in sequential data assimilation.

\end{abstract}

\newpage

\section{Introduction} 

The problem of estimating the state of an evolving system from an incomplete set of noisy observations is the central theme of the classical estimation theory and it constitutes the object of the filtering procedures (Jazwinski, 1970). 
The filtering theory finds its natural applications in experimental and applied physics in general as well as in engineering, where the aim is to extract as much information as possible about a natural or laboratory scale phenomenon on the basis of limited amount of noisy observations.  

The classical filtering and prediction problem has encountered a natural field of application in the context of numerical weather and oceanic prediction, where it is usually referred to as {\it data assimilation} (Bengtsson et al., 1981; Daley, 1991; Kalnay, 2003). The ultimate goal in this context is to provide the best possible estimate of an unknown system's state by using all the information available (Talagrand, 1997). Typically, it is based on the information provided by the laws governing the unknown system ({\it i.e.} the model) under the form of a continuous dynamical system, while the observations are available only at discrete times.
In sequential assimilation the system's state estimate, given by a solution of the model equations, is updated at each time when observations are available. This update is usually referred to as the {\it analysis}. 
The procedure consists therefore of a sequence of analyses performed at observation times and of integrations of the model between successive analyses (Talagrand, 1997).

It is well known that in the case of a linear dynamics, and a linear relation between observations and the system's state variables, the filtering problem can be expressed via the Kalman filter (KF) equations (Kalman, 1960; Jazwinski, 1970). 
The KF formulation is an elegant and comprehensive mathematical description: a closed set of equations providing the optimal linear solution of the filtering problem, that is to say the state estimate and the associated error covariances, in the hypothesis that the observation and the model error are Gaussian, white-in-time, and mutually uncorrelated. 

For nonlinear dynamics, an extension of the KF formulation has been proposed, the extended Kalman filter (EKF) (Jazwinski, 1970), which has been largely studied in geophysical contexts (see {\it e.g.}, Ghil, 1989; Ghil and Malanotte, 1991; Miller et al., 1994). In the EKF the system state evolves according to the full nonlinear dynamics, while the associated error covariance are propagated in time through the linearized dynamics. As long as the dynamics of the estimation error is well approximated by the tangent linear equations, the EKF has proven to be quite efficient (Miller et al, 1994; Yang et al., 2006) while preserving one of the most attractive feature of the KF equations, namely the time propagation of the error covariance, which is desirable especially in the case of chaotic dynamics where the error evolution is conditioned by the local instabilities of the flow. 

Besides the computational problems and the intrinsic inaccuracy related to the use of a linearized model for propagating the error statistics, another fundamental issue in the applications of KF-like sequential assimilations is the assumption on the model error. 
Indeed, models used for meteorological and climate prediction still have large and unknown deficiencies, including inaccurate parameterizations of physical processes not accounted for in the model as well as error in the numerical integration scheme. In this circumstance, prediction errors are essentially due to the use of inadequate models and their chaotic dynamics leads to the rapid amplification of these errors as well as of initial condition ones. Modern, advanced, data assimilation techniques have tackled this latter problem so that progressively more accurate initial condition for environmental prediction are made available (for a comprehensive description of data assimilation development see Daley, 1991; Kalnay, 2003). 
Even though it is still not possible to eliminate model deficiencies, a number of solutions have been proposed in recent years to estimate and/or account for model error in data assimilation. 
The state augmentation method (Jazwinski, 1970) was primarily introduced in the context of Kalman filtering. In this method, the state estimation problem is formulated in terms of an augmented state vector which includes, along with the state estimate, a set of parameters used for the model error representation. This approach has been applied successfully in both KF-like and variational assimilation (see {\it e.g.} Zupanski, 1997; Nichols, 2003; Zupanski and Zupanski, 2006). Dee and da Silva (1998) proposed an algorithm to estimate and remove the biases in the background field in a data assimilation system due to additive systematic model error. The method was implemented for the bias correction of the humidity analysis component of the Goddard Earth Observing System (GEOS) assimilation system (Dee and Todling, 2000) and more recently in the context of the European Center for Medium Weather Forecast (ECMWF) ocean data assimilation (Balmaseda et al., 2007).

A key ingredient of the state augmentation technique is the definition of a "model" for the model error (see {\it e.g.} Nichols, 2003). Recently Zupanski and Zupanski (2007) have described the model error evolution using a first order Markov process. 
A similar assumption was already used in Daley (1992) to investigate the impact of time correlated model errors in Kalman filtering. He showed the detrimental effect on the analysis accuracy of the absence of time correlation in the description of the model error evolution. More generally these studies have highlighted the urgent requirement of a deeper understanding of model error dynamics.  

In recent years, a significant body of work on model error dynamics appeared, in particular in relation to predictability studies (see {\it e.g.} Reynolds et al., 1994; Orrel et al., 2001; Vannitsem and Toth, 2002).   
In particular, the foundations of the dynamics of deterministic model errors have been laid in Nicolis (2003) and Nicolis (2004). In these works, some generic features of model error dynamics such as the existence of an universal quadratic law for the short time mean square error evolution, whose extent of validity is related to the Lyapunov spectrum of the underlying dynamics, have been highlighted. The relative roles of deterministic dynamics and stochastic error sources terms have also been investigated (Nicolis, 2004).

These results are the starting point of the present study whose ultimate goal is to investigate if these generic features found on the deterministic model error dynamics can be efficiently incorporated in the EKF. The purpose here is twofold. First, we investigate the dynamics of state estimation error when both initial condition and model errors are present, and second, an EKF formulation which takes into account the model error dynamics is examined. 
These questions are analyzed in the context of a low order chaotic system (Lorenz, 1996) in which model error originates from an inaccurate specification of some of the model parameters. 

The paper is organized as follows. In Sect. 2 the formulation of the problem is presented and deterministic evolution equations for the mean and covariance error are derived, along with suitable short time approximations. The results are used to describe model error evolution within the analysis intervals and lead to a formulation of the EKF in which the deterministic character of the model error is accounted for (Sect. 3). Section 4 contains the results of the numerical analysis; the first part describes the experiments related to the analytical derivation of Sect. 2 on the estimation error evolution, while in the second, the EKF with the deterministic model error description is tested and compared with the white noise model error formulation. Conclusions are drawn in Sect. 5. 

\vspace{8pt}

\section{\label{Err-Dyn}  Error dynamics in the presence of initial condition and model errors}

In this section a general equation for the evolution of the error associated to the estimate of the state of a system is derived, based on the use of an imperfect model of the system's dynamics, and an approximate knowledge of its initial state.

Let the (unknown) "true" dynamics, the {\it nature}, be represented in the form:
\begin{equation}
\label{truth}
\frac{d{\bf y}(t)}{dt} = {\bf f} ({\bf y}(t),{\bf \lambda})
\end{equation}
The state vector, ${\bf y}$, lies in an $I$-dimensional vector space, the governing law ${\bf f}$, typically a nonlinear function, defined in ${\cal R}^I$ and ${\bf \lambda}$ is a $P$-dimensional vector of parameters. The function ${\bf f}$ can have an explicit dependence on time but it is dropped here to simplify the notation. 

Let ${\bf x}$ be the variable of the {\it model} at our disposal obeying: 
\begin{equation}
\label{model}
\frac{d{\bf x}(t)}{dt} = {\bf g} ({\bf x}(t),{\bf \lambda}^{'}),
\end{equation}
where ${\bf g}$ is a nonlinear function. Typically, ${\bf g}$ is defined on a phase space which is different from ${\bf f}$ in Eq. (\ref{truth}), and ${\bf \lambda}^{'}$ is a $Q$-dimensional vector ($Q\neq P$).

The error in the estimate of ${\bf y}$, referred to as {\it model error}, originated by the approximate description of the true dynamics, Eq. (\ref{model}). Different sources of model error are possible such as, for instance, errors related to an inadequate description of some physical processes and/or to the limited representation of relevant scales. In this study, we focus on the situation in which the model and the true trajectories span the same phase space. Model error is due only to uncertainties in the specification of the parameters appearing in the evolution law ${\bf g}$. This formulation accounts, for instance, for errors in the description of some physical processes (dissipations, external forcing, etc.) represented by the parameters. The role of the unresolved scales in the model dynamics and the consequence for the data assimilation will be addressed in a future work.

The model state and parameter, ${\bf x}$ and $\lambda^{'}$, are therefore $I$-dimensional and $P$-dimensional vectors respectively, and the evolution equation (\ref{model}) can be rewritten as:  
\begin{equation}
\label{dyn-model}
\frac{d{\bf x}}{dt} = {\bf f}({\bf x},\lambda^{'})  
\end{equation}

An equation for the evolution of the state estimation error $\delta{\bf x}(t)={\bf y}(t)-{\bf x}(t)$ can be obtained by taking the difference between Eqs. (\ref{truth}) and (\ref{dyn-model}). The evolution of $\delta{\bf x}(t)$ depends on the error estimate at the initial time $t=t_0$ (initial condition error $\delta{\bf x}(t_0)=\delta{\bf x}_0$) and on the model error. If $\delta{\bf x}$ is "small", the linearized dynamics provides a reliable approximation of the actual error evolution. The linearization is made along a model trajectory, solution of Eq. (\ref{dyn-model}), by expanding, to the first order in $\delta{\bf x}$ and $\delta\lambda (=\lambda-\lambda^{'})$, the difference between Eqs. (\ref{truth}) and (\ref{dyn-model}):
\begin{equation}
\label{dyn-err-deriv}
\dfrac{d\delta{\bf x}}{dt} \approx \frac{\partial{\bf f}}{\partial{\bf x}}|_{{\bf x}(t)}\delta{\bf x} + \frac{\partial{\bf f}}{\partial\lambda}|_{\lambda^{'}}\delta\lambda   
\end{equation}
The first partial derivative on the rhs of Eq. (\ref{dyn-err-deriv}) is the Jacobian of the model dynamics evaluated along its trajectory. The second term, which corresponds to the model error, will be denoted $\delta{\bf \mu}$ hereafter to simplify the notation; $\delta{\bf \mu}=\frac{\partial{\bf f}}{\partial\lambda}|_{\bf\lambda^{'}}\delta{\bf \lambda}$. 

The solution of Eq. (\ref{dyn-err-deriv}), with initial condition $\delta{\bf x}_0$ at $t=t_0$, reads:
\begin{equation*}
\delta{\bf x}(t) \approx {\bf M}_{t,t_0}\delta{\bf x}_0 + \int^{t}_{t_{0}} d\tau{\bf M}_{t,\tau}\delta{\bf \mu}(\tau)  
\end{equation*}
\begin{equation}
\label{dyn-err-solution}
= \delta{\bf x}^{ic}(t) + \delta{\bf x}^{m}(t) 
\end{equation}
with ${\bf M}_{t,t_0}$ being the fundamental matrix (the propagator) relative to the linearized dynamics along the trajectory between $t_0$ and $t$. We point out that $\delta{\bf \mu}$ and ${\bf M}_{t,\tau}$ in (\ref{dyn-err-solution}) depend on $\tau$ through the state variable ${\bf x}$. Equation (\ref{dyn-err-solution}) states that, in the linear approximation, the error in the state estimate is given by the sum of two terms, one relative to the evolution of initial condition error, $\delta{\bf x}^{ic}$, and another one relative to the model error, $\delta{\bf x}^{m}$. 
The presence of the fundamental matrix ${\bf M}$ in the expression for $\delta{\bf x}^{m}$ suggests that similarly to what occurs for the initial condition error, the instabilities of the flow play a role in the dynamics of model error as well. 

Let us now apply to Eq. (\ref{dyn-err-solution}) the expectation operator defined locally around the reference trajectory, by sampling over an ensemble of initial conditions and model errors:
\begin{equation}
\label{exp-oper}
< . > = \int d(\delta{\bf x}_{0})\rho_0(\delta{\bf x}_{0})\int d(\delta{\bf\mu})\rho_m(\delta{\bf\mu})
\end{equation}
where $\rho_0$ and $\rho_m$ are the probability density function for the initial condition and the probability density of the model error respectively.
This definition is consistent with the expectation operator usually used in state estimation algorithms.
We then get the evolution equation for the mean estimation error along a reference trajectory:    
\begin{equation*}
<\delta{\bf x}(t)> \approx {\bf M}_{t,t_0}<\delta{\bf x}_0> + \int^{t}_{t_{0}} d\tau{\bf M}_{t,\tau}<\delta{\bf \mu}(\tau)>   
\end{equation*}
\begin{equation}
\label{dyn-err-solution-mean}
= <\delta{\bf x}^{ic}> + <\delta{\bf x}^m> 
\end{equation}

In a perfect model scenario an unbiased state estimate at time $t_0$ ($<\delta{\bf x}_0>=0$) will evolve, under the linearized dynamics, into an unbiased estimate at time $t$. This is not necessarily true when model error is present and, depending on the properties of model error, an initially unbiased estimate can evolve into a biased one. The important factor controlling the evolution of the mean state estimation error is the model error mean $<\delta{\bf \mu}(t)>$. In view of the hypothesis made above on the type of model error, the latter is expressible as a function of the model variables and of the parametric error $\delta{\bf \lambda}$, so that in the general case we expect $<\delta{\bf \mu}(t)>$ to be different from zero. This fact has important implications for classical least-square (or Bayesian) based data assimilation algorithms which are derived assuming that the errors associated to each piece of information entering the analysis update are Gaussian and unbiased (Talagrand, 1997). In that context, if the bias is not properly accounted for and removed, the resulting analysis state will be biased (Dee and Da Silva, 1998). Equation (\ref{dyn-err-solution-mean}) provides a way to estimate the bias due to an incorrect specification of the model parameters. 

Formal expressions for other moments of the unknown error probability density function (PDF) can be derived similarly. Since the focus here is on data assimilation related problems, we will make the simplifying assumption that the underlying error PDF is (or is close to) Gaussian. As a consequence the full PDF is mainly characterizable by only its first and second moments: the mean and the covariance. 

The evolution equation of the state estimation error covariance matrix can be obtained by taking the expectation of the external product of $\delta{\bf x}(t)$ by itself. As above, the expectation is made over a large sample of initial conditions and model errors, assuming that the estimation error biasi is known and removed from the background error field, and leads to:
\begin{equation*}
{\bf P}(t) = < (\delta{\bf x}(t))(\delta{\bf x})^T(t) >  
\end{equation*}
\begin{equation}
\label{dyn-cov-general}
\approx  {\bf P}^{ic}(t) + {\bf P}^{m}(t) + {\bf P}^{corr}(t) + ({\bf P}^{corr})^T(t)
\end{equation}
where:
\begin{equation}
\label{PIC}
{\bf P}^{ic}(t) = {\bf M}_{t,t_0} < (\delta{\bf x}_0)(\delta{\bf x}_0)^T > {\bf M}^T_{t,t_0} 
\end{equation}
\begin{equation}
\label{PMOD}
{\bf P}^{m}(t) =  \int^{t}_{t_{0}} d\tau \int^{t}_{t_{0}} d\tau^{'} {\bf M}_{t,\tau}< (\delta{\bf \mu}(\tau))(\delta{\bf \mu}(\tau^{'}))^T)> {\bf M}^T_{t,\tau^{'}}
\end{equation}
\begin{equation}
\label{PCORR}
{\bf P}^{corr}(t) = {\bf M}_{t,t_0}<(\delta{\bf x}_0)\left( \int^{t}_{t_{0}}d\tau{\bf M}_{t,\tau}\delta{\bf \mu}(\tau)\right)^T> 
\end{equation}
The four terms of the r.h.s. of Eq. (\ref{dyn-cov-general}) depict the evolution of the initial condition error covariance, the model error covariance and their cross correlation matrices, respectively. The evolution of the quadratic error is obtained by taking the trace of the corresponding covariance matrix. It is interesting to note that the net effect of the correlation between initial condition and model error may result in a reduction of the total estimation error. 

We now turn briefly to the situation in which the model error, $\delta{\bf \mu}$, is assumed to be an additive random disturbance. In this case, Eq. (\ref{dyn-err-deriv}) takes the form of a stochastic differential equation whose solution depends on the property of the random process itself. 
For white noise Gaussian process $<\delta{\bf \mu}>=0$ and $<(\delta{\bf\mu}(t))(\delta{\bf\mu}(t^{'}))^T>={\bf Q}\delta(t-t^{'})$, where ${\bf Q}=<(\delta{\bf\mu}(t))(\delta{\bf\mu}(t))^T>$ is a positive definite matrix representing the covariance of the process and $\delta(t-t^{'})$ is a Dirac delta function.

In analogy to the case of deterministic model error, we can derive an expression for the evolution of estimation error covariance matrix when the model error is treated as a white noise process, by taking the expectation value of the external product of Eq. (\ref{dyn-err-solution}) by itself:
\begin{equation}
\label{PWN}
{\bf P}_{wn}(t) =  {\bf P}^{ic}(t) + {\bf P}^{m}_{wn}(t)  
\end{equation}
with ${\bf P}^{ic}$ the same as in Eq. (\ref{PIC}), and ${\bf P}^{m}_{wn}$ representing the model error covariance matrix in this white noise case:
\begin{equation*}
{\bf P}^{m}_{wn}(t) = \int^{t}_{t_{0}} \int^{t}_{t_{0}} d\tau d\tau^{'}{\bf M}_{t,\tau} < \delta{\bf\mu}(\tau) \delta{\bf \mu}(\tau^{'})^T)>{\bf M}_{t,\tau^{'}}^T    
\end{equation*}
\begin{equation}
\label{PMWN}
= \int^{t}_{t_{0}}{\bf M}_{t,\tau}{\bf Q}{\bf M}_{t,\tau}^T d\tau   
\end{equation}
with the subscript $"wn"$ standing for "white noise". 

We finally point out that the mean and covariance matrix evolution equations derived so far, are based on the expectation operator defined in Eq. (\ref{exp-oper}). This means that we implicitly assume to have a reference trajectory, and to be interested in exploring the distribution of possible errors in the estimation of the trajectory itself. This is typically the case in data assimilation applications when one is usually interested in evaluating the error associated to the estimate of a given trajectory, solution of a model integration. The same averaging procedure can then be repeated by sampling over an ensemble of states over the system's attractor (through the probability density function associated with the model attractor) in order to get information independent of the initial conditions as usually done in traditional analysis of error dynamics (Nicolis, 1992).   \\ 

{\bf Short time error evolution}\\

Let us now turn to a short-time approximation of Eqs. (\ref{dyn-err-solution}), (\ref{dyn-cov-general}) and (\ref{PMWN}).

We proceed by expanding Eq. (\ref{dyn-err-solution}) in Taylor series, up to the first non trivial order, only for the model error term $\delta{\bf x}^m$ while keeping the initial condition term, $\delta{\bf x}^{ic}$, unchanged.  

In this case, the model error $\delta{\bf x}^{m}$ evolves linearly with time according to:
\begin{equation}
\label{dyn-err-mod_Ib}
\delta{\bf x}^{m} \approx \delta{\bf \mu}_0(t-t_0) 
\end{equation}
where $\delta{\bf \mu}(t_0)=\delta{\bf \mu}_0$.

By adding the initial condition error term $\delta{\bf x}^{ic}$, we get a short time approximation of Eq. (\ref{dyn-err-solution}):
\begin{equation}
\label{dyn-err-der_Ib}
\delta{\bf x}(t) \approx {\bf M}_{t,t_0}\delta{\bf x}_0 + \delta{\bf \mu}_0(t-t_0) 
\end{equation}
For the mean error we get:
\begin{equation}
\label{errore-medio}
<\delta{\bf x}(t)> \approx {\bf M}_{t,t_0}<\delta{\bf x}_0> + <\delta{\bf \mu}_0>(t-t_0) 
\end{equation}
As already noted by Nicolis (2003), the mean model error evolves linearly in time as long as the average $<\delta{\bf \mu}_0>$ is different from zero, otherwise the evolution is conditioned by higher orders of the Taylor expansion. 

We remark that the two terms in the short time error evolution Eqs. (\ref{dyn-err-der_Ib}) and (\ref{errore-medio}), are not on equal footing since, in contrast to the model error term, which has been expanded up to the first nontrivial order in time, the first r.h.s. term (the term describing the initial condition error evolution) contains all the orders of times (t,t$^2$,...,t$^n$). The point is that, as explained in the sequel, we intend to use these equations to model the error evolution in conjunction with the technique of data assimilation. As discussed in Section 3 in this technique it is customary to proceed with the full matrix ${\bf M}$ as it appears in Eqs. (\ref{dyn-err-der_Ib}) and (\ref{errore-medio}). We stress however that since the assimilation intervals of a data assimilation cycle are short effectively only the first terms of a Taylor expansion of ${\bf M}$ are relevant. This secures the overall consistency of our evaluation. 

Taking the expectation value of the external product of Eq. (\ref{dyn-err-der_Ib}) by itself, we get:
\begin{equation*}
{\bf P}(t) \approx {\bf M}_{t,t_0} < (\delta{\bf x}_0)(\delta{\bf x}_0)^T > {\bf M}_{t,t_0}^T  + \\
\end{equation*}
\begin{equation*}
+ \left[ < (\delta{\bf \mu}_0)(\delta{\bf x}_0)^T >{\bf M}^T_{t,t_0} + {\bf M}_{t,t_0} < (\delta{\bf x}_0)(\delta{\bf \mu}_0)^T > \right](t-t_0)  \\
\end{equation*}
\begin{equation}
\label{P-approx}
+  < (\delta{\bf \mu}_0)(\delta{\bf \mu}_0)^T > (t-t_0)^2
\end{equation}

Equation (\ref{P-approx}) is the short time evolution equation, in this linearized setting, for the error covariance matrix in the presence of both initial condition and model errors. 
Note that while model error is bound to evolve quadratically, in agreement with the results of Nicolis (2003), the correlation errors behave linearly with time and these terms may have a compensating effect resulting in a reduction of the total error. 
The extent of validity of the quadratic short time regime of the mean square error is related to the largest (in absolute value) exponent (the most negative one for the large class of dissipative systems), see Nicolis (2003). This result is also confirmed by the numerical results presented below. 

A case which can have practical relevance, is obtained when the model error is independent of $\delta{\bf x}_0$ (error in the specification of external source parameters, for instance) and of unbiased initial condition errors. In these circumstances Eq. (\ref{P-approx}) reads: 
\begin{equation*}
{\bf P}(t) \approx {\bf M}_{t,t_0}< (\delta{\bf x}_0)(\delta{\bf x}_0)^T > {\bf M}^T_{t,t_0}   
\end{equation*}
\begin{equation}
\label{dyn-cov-approx-q-nocorr}
+ < (\delta {\bf \mu}_0)(\delta {\bf \mu}_0)^T >(t-t_0)^2 
\end{equation}
and the correlation terms cancel. 

A similar relation can be built when considering a white noise model error term. Developing the second term of Eq. (\ref{PWN}) in Taylor series and keeping the dominant order alone one gets:
\begin{equation}
\label{PWNapprox} 
{\bf P}(t) \approx {\bf M}_{t,t_0}< (\delta{\bf x}_0)(\delta{\bf x}_0)^T > {\bf M}^T_{t,t_0} + {\bf Q}(t-t_0)
\end{equation}

By comparing Eqs. (\ref{dyn-cov-approx-q-nocorr}) and (\ref{PWNapprox}) we conclude that while the model error covariance matrix evolves linearly with time if the model error acts as a white noise process, it is bound to evolve quadratically, for short times, in the case of deterministic model error.

Finally, as discussed previously, the average procedure at the basis of Eqs. (\ref{errore-medio}) and (\ref{dyn-cov-approx-q-nocorr}), can also be performed by averaging over an ensemble of states on the system's attractor (symbolically represented by $<< . >>$), leading to:
\begin{equation*}
<<\delta{\bf x}(t)>> \approx <<{\bf M}_{t,t_0}\delta{\bf x}_0>> 
\end{equation*}
\begin{equation}
\label{errore-medio-attr}
+ <<\delta{\bf \mu}_0>>(t-t_0)
\end{equation}
\begin{equation*}
{\bf P}(t) \approx <<({\bf M}_{t,t_0} \delta{\bf x}_0)({\bf M}_{t,t_0} \delta{\bf x}_0)^T >>  
\end{equation*}
\begin{equation}
\label{dyn-cov-approx-q-nocorr-attr}
+ << (\delta {\bf \mu}_0)(\delta {\bf \mu}_0)^T >>(t-t_0)^2
\end{equation}

Relations (\ref{errore-medio-attr}) and (\ref{dyn-cov-approx-q-nocorr-attr}) are used in Sect. 4.

\section{\label{Analysis} Extended Kalman filter in the presence of model error}

In this section we first revise the classical EKF equations, then a formulation of the filter in which the deterministic dynamics of the parametric model error is accounted for is presented.

Let assume that a set of $M<I$ noisy observations of the true system (\ref{truth}), stored as the components of an $M$-dimensional observation vector ${\bf y}^o$, is available at the regularly spaced discrete times $t_k=t_0+k\tau$, $k=1,2...$, with $\tau$ being the assimilation interval; that is:
\begin{equation}
\label{obs}
{\bf y}^o_k={\cal H}({\bf y}_k)+\epsilon_k
\end{equation}
where $\epsilon$ is the observation error, assumed to be Gaussian with known covariance matrix ${\bf R}$ and uncorrelated in time. ${\cal H}$ is the (possibly nonlinear) observation operator which maps from model to observation space ({\it i.e.} from model to observed variables) and may involve spatial interpolations (or spectral to physical space transformation in the case of spectral models) as well as transformations based on physical laws for indirect measurements (Kalnay, 2003).
 
It is convenient to write the model equations as a discrete mapping from time $t_k$ to $t_{k+1}$:
\begin{equation}
\label{model-dyn-discrete}
{\bf x}^{f}_{k+1} = {\cal M}{\bf x}^{a}_k
\end{equation}
${\bf x}^{f}$ and ${\bf x}^{a}$ being the forecast and analysis states respectively, ${\cal M}$ the nonlinear model forward operator (the resolvent of Eq. (\ref{dyn-model})). 

For the EKF, as well as for most least-square based assimilation schemes, the analysis update equation is (Jazwinski, 1970; Daley, 1991):
\begin{equation}
\label{analysis-update}
{\bf x}^{a}_k=\left[{\bf I} -{\bf K}_k{\cal H}_k\right]{\bf x}^{f}_k + {\bf K}_k{\bf y}^{o}_k={\bf x}^{f}_k+{\bf K}_k{\bf d}_k  
\end{equation}
where ${\bf d}={\bf y}^{o}-{\cal H}{\bf x}^{f}$ is the $M$-dimensional vector of {\it innovation}. The {\it gain} matrix ${\bf K}$ is given by:
\begin{equation}
\label{gain matrix}
{\bf K}={\bf P}^{f}{\bf H}^T\left[{\bf H}{\bf P}^{f}{\bf H}^T + {\bf R}\right]^{-1}  
\end{equation}
where ${\bf P}^{f}$ is the $I\times I$ forecast error covariance matrix and ${\bf H}$ the linearized observation operator ({\i.e.} a $M\times I$ real matrix); we use the unified notation given in Ide et al. (1997) and the temporal index $k$ has been omitted in Eq. (\ref{gain matrix}) for clarity. The essence of an assimilation scheme is embedded in the gain matrix: it characterizes the algorithm, the way it "uses" the observations. The characteristics of the observational network are described through the observation operator ${\cal H}$. 

In the EKF, the forecast error covariance matrix, ${\bf P}^f$, is obtained by linearizing the model around its trajectory between two successive analysis times $t_{k}$ and $t_{k+1}$. Similarly to what has been already done in relation with Eq. (\ref{dyn-err-solution}), the forecast error at time $t_{k+1}$, $\delta{\bf x}^f_{k+1}$, can be approximated as:
\begin{equation}
\label{fc_err}
\delta{\bf x}^f_{k+1} = {\bf M}_{k+1,k}\delta{\bf x}^a_k + \delta{\bf x}^{m}_{k+1}
\end{equation}  
The model error term $\delta{\bf x}^{m}_{k+1}$ can be thought to represent the sum of all the contributions to the state vector which are not accounted for by the tangent linear propagator. 

Assuming that the model error is uncorrelated with the analysis error, the evolution equation of the forecast error covariance matrix within the assimilation interval is given by: 
\begin{equation}
\label{Pf_EKF}
{\bf P}^f_{k+1} = {\bf M}_{k+1,k}{\bf P}^a_k{\bf M}^T_{k+1,k} + {\bf P}^{m}_{k+1}
\end{equation}  
The analysis and forecast error covariance matrices are related through the gain matrix, according to:
\begin{equation}
\label{Pa_EKF}
{\bf P}^a_k = [{\bf I}-{\bf K}_k{\bf H}_k]{\bf P}^f_k
\end{equation}  
Expression (\ref{model-dyn-discrete}) and (\ref{Pf_EKF}) are usually referred to as the EKF prediction equations, while Eqs. (\ref{analysis-update}), (\ref{gain matrix}) and (\ref{Pa_EKF}) as the EKF analysis equations. If all hypotheses are verified, the analysis given by Eq. (\ref{analysis-update}) corresponds to the minimum variance, unbiased, estimate of the true system's state. 

According to the discussion in Section 2, if the model error is an additive white noise, its impact on the estimation error is related to the covariance of the random process, the matrix ${\bf Q}$. Consequently, assuming to have access to (or to an estimate of) ${\bf Q}$, the model error covariance matrix in Eq. (\ref{Pf_EKF}) can be estimated through the short time linear approximation, Eq. (\ref{PWNapprox}), as ${\bf P}^m={\bf Q}\tau$. 

On the other hand, the short time evolution of the deterministic model error covariance is bound to be quadratic. In this latter case, by neglecting the correlation terms, the model error covariance matrix can be approximated as ${\bf P}^m={\bf Q}\tau^{2}$, where ${\bf Q}$ is the covariance of the uncertainties associated to the deterministic model error. Thus, the two estimates of ${\bf P}^m$ based respectively on the white noise or on the deterministic process assumption, differ by a multiplicative factor equal to the assimilation interval $\tau$. 

Given these premises, we investigate the possibility of using the short time approximation, Eq. (\ref{dyn-cov-approx-q-nocorr}), to describe the error evolution between assimilation times along the EKF analysis cycle. As long as the observational forcing is frequent enough (small $\tau$) and the error is efficiently reduced by the assimilation of observations, the short time error dynamics should provide a reliable description of the actual model error evolution between two successive analysis. Similarly, Eq. (\ref{errore-medio}) may be used in an EKF analysis cycle to estimate the bias in the forecast error; this bias can then be removed from the forecast field before the latter is used as the background in the analysis update, Eq. (\ref{analysis-update}).

Let us suppose to have access to a statistical information about the parametric error of our model, under the form of the mean $<\delta{\bf\mu}>$ and its covariance ${\bf Q}=< (\delta{\bf\mu} - <\delta{\bf\mu}>)(\delta{\bf\mu} - <\delta{\bf\mu}>)^T >  $.
At the analysis times, the forecast bias due to the model error can be estimated according to the short time approximation, Eq. (\ref{errore-medio}), that is: 
\begin{equation}
\label{bias_es}
<\delta{\bf x}^m>=<\delta{\bf\mu}>(t_{k+1}-t_k)=<\delta{\bf\mu}>\tau
\end{equation}
The bias can then be removed from the forecast field before the EKF analysis is performed: 
\begin{equation}
\label{back_rm}
{\tilde{\bf x}}^f_k={\bf x}^f_k-<\delta{\bf x}^m>
\end{equation}
The new background state ${\tilde{\bf x}}^f_k$ is finally used for the analysis update, Eq. (\ref{analysis-update}).

Similarly, the model error covariance matrix can be estimated with the constant matrix:
\begin{equation}
\label{PMapprox}
{\bf P}^m={\bf P}^m_{dp} \approx <(\delta{\bf\mu}-<\delta{\bf\mu}>)(\delta{\bf\mu}-<\delta{\bf\mu}>)^T>\tau^2
\end{equation}
where the suffix "dp" stands for deterministic process. Or alternatively in the case of the white noise model error assumption, 
\begin{equation}
\label{PMWNapprox}
{\bf P}^m={\bf P}^m_{wn}\approx <(\delta{\bf\mu}-<\delta{\bf\mu}>)(\delta{\bf\mu}-<\delta{\bf\mu}>)^T>\tau
\end{equation}
These formulations of the EKF in which model error is treated as a deterministic process or a white noise are tested numerically through the implementation of Observation System Simulation Experiments (OSSE) (Bengtsson et al., 1981), see Section 4.2. 

In both formulations the bias is removed, Eqs. (\ref{bias_es}) and (\ref{back_rm}). Strictly speaking, if the model error is a white noise, its mean is identically zero, and no model related bias should be present. In this case, the matrix ${\bf P}^m_{wn}$ can be interpreted to only account for the variability of the model error around its mean.

Note that in practice, the statistical information on the model error, $<\delta\bf\mu>$ and ${\bf Q}$, are not easy to evaluate. In the present work, since the origin of this model error is known, we perform the statistics over the whole attractor in order to get an invariant estimate of the model error moments.  

\section{Numerical experiments}

In the first part of this Section we study the evolution of the estimation error in the presence of errors in the initial conditions and in the model parameters. We focus on the investigation of the accuracy of the short time error and error covariance dynamics, Eqs. (\ref{errore-medio}) and (\ref{P-approx}) respectively. In the second part of the section, OSSEs are performed with the aim of testing the formulation of the EKF in which the dynamics of the model error is accounted for.

These questions are investigated in the context of a low-order atmospheric model giving rise to chaotic dynamics. The model, introduced by Lorenz (1996), possesses 36 scalar variables representing the values of some meteorological quantity $x_i$, $i=\{1,...,36\}$, along a latitudinal circle. The evolution equations read:
\begin{equation*}
\frac{dx_i}{dt}=\alpha(x_{i+1}-x_{i-2})x_{i-1} - \beta x_i + F 
\end{equation*}
\begin{equation}
\label{model-expr}
= f({\bf x},\alpha,\beta, F), \qquad  i=\{1,...,36\}
\end{equation}
The quadratic terms simulate the advection, the linear term the internal dissipation, with ${\bf\alpha}$ and ${\bf\beta}$ being the advection and dissipation parameters respectively, while $F$ represents the external forcing; a detailed description of the model can be found in Lorenz (1996). The numerical integrations have been performed using a fourth-order Runge-Kutta scheme with a time step of 0.0083 units, corresponding to 1 hour of simulated time. 

For ${\bf\alpha}={\bf\beta}=1$ and $ F=8$, the model behaves chaotically and the first Lyapunov exponent, computed for a period of 200 years after a spin-up of $10$ years, is equal to $\sigma_{1}=0.33$ (day)$^{-1}$ and corresponds to a doubling time of about 2 days. Reducing the external forcing, the system becomes progressively more stable (for sufficiently small $ F$ all solutions decay to $x_1=...=x_{36}=F$); the opposite occurs when the external forcing is increased. By increasing/decreasing the internal dissipation the system becomes more stable/unstable as expected. When the advection coefficient is modified the stability properties do not vary significantly. The asymptotic stability properties of Eq. (\ref{model-expr}), for a set of parameters values are summarized in Tab. 1. This set of parameters values are used in the numerical experiments, while the true dynamics is represented by the model with the canonical values ${\bf\alpha}={\bf\beta}=1$ and $ F=8$. The amplitude of the parameter error ($20\%$) has been chosen as the minimum value which lead to sensible deterioration of the quality of the model prediction.   
The focus is placed on the two model configurations given in the last two lines of Tab. 1 (hereafter referred to as $"C_I"$ and $"C_{II}"$, respectively), since in these cases all the three model parameters are modified simultaneously.

According to Eq. (\ref{dyn-err-deriv}), the state dependent model error can be written as:
\begin{equation*}
\delta \mu_i({\bf x},\Delta\lambda) = \Delta\alpha(x_{i+1}-x_{i-2})x_{i-1} - \Delta\beta x_i + \Delta F 
\end{equation*}
\begin{equation}
\label{model_error}
i=\{1,...,36\}
\end{equation}
where the terms $\Delta\alpha$, $\Delta\beta$ and $\Delta F$ are the errors in the specification of the advection, dissipation and external forcing respectively and $\Delta\lambda=\lambda-\lambda^{'}=(\Delta\alpha,\Delta\beta,\Delta F)$, and the true parameters are $\lambda=(\alpha,\beta,F)=(1,1,8)$.

\subsection{Error dynamics in the presence of parametric model error} 

The experiments described in this section are designed to investigate the accuracy of the linearized, short time approximations, Eqs. (\ref{errore-medio-attr}) and (\ref{dyn-cov-approx-q-nocorr-attr}). As stated above, the true dynamics is given by the model, Eq. (\ref{model-expr}), with ${\bf\alpha}={\bf\beta}=1$ and $F=8$. We study the quality of the state estimate prediction, when the approximate model is described with the parameters of Tab. 1.

A sample of $10^5$ initial conditions over the true system attractor is generated through a long time integration. A set of $10^5$ model initial conditions are produced such that the initial error is $\delta{\bf x}_0 \in {\cal N} (0,\sigma_{0})$; $\sigma_0$ is fixed to $10\%$ of the system's natural variability. 

\begin{figure*}
\centering
\includegraphics{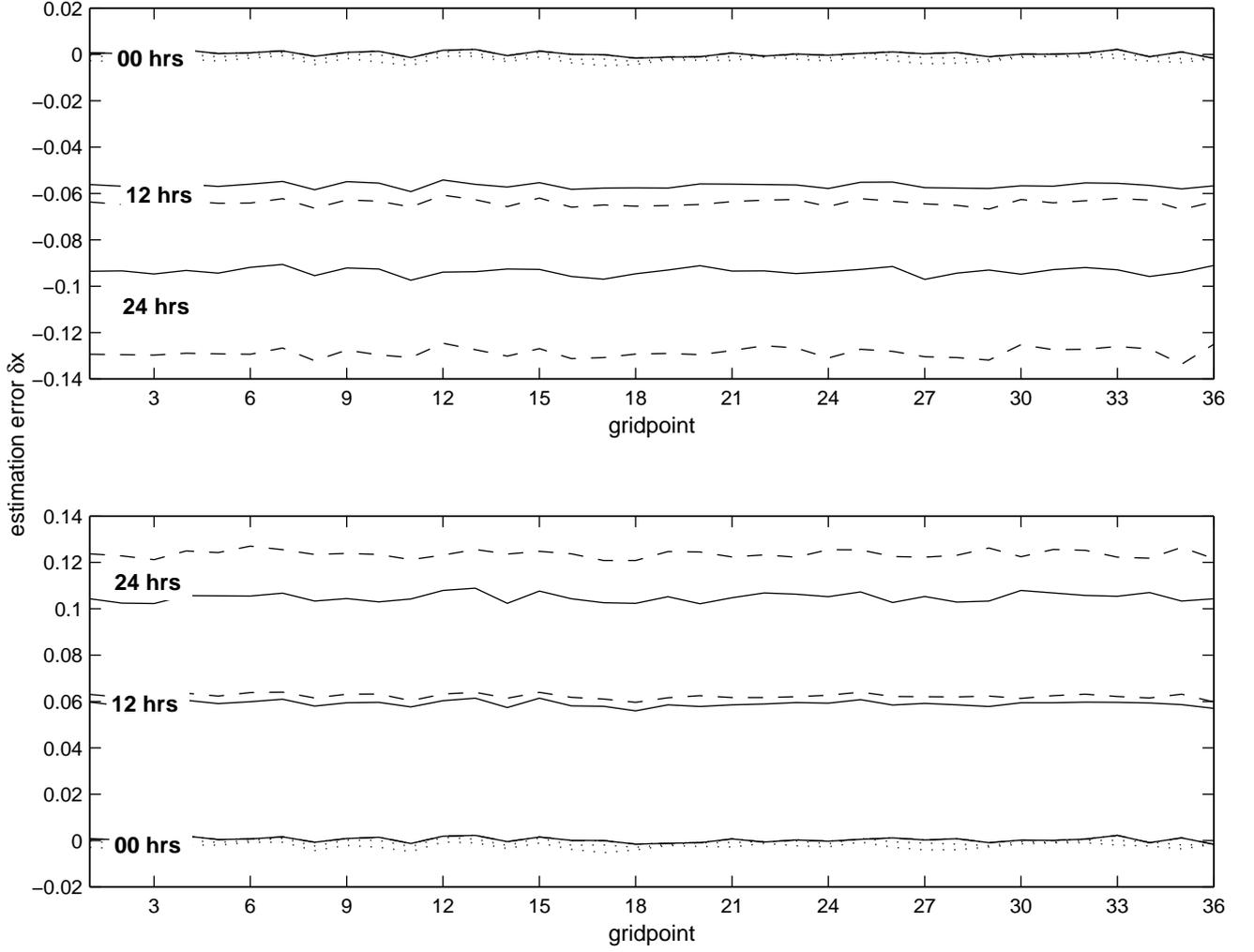}
\caption{\label{FIG1} Mean error evolution for configurations $C_I$ (top panel) and $C_{II}$ (bottom panel). Actual error (continuous line); linearly evolved mean initial condition error, $<{\bf M}_{t,t_0}\delta{\bf x}_0>$, (dotted line); total error, model plus initial condition, approximated by Eq. (\ref{errore-medio-attr}) (dashed line). Error is shown as a function of the model gridpoints at times $t=0, 12, 24$ hours.}
\end{figure*}

In Fig. \ref{FIG1} the mean error evolution is compared with the short time approximations, Eqs. (\ref{errore-medio-attr}) and (\ref{dyn-cov-approx-q-nocorr-attr}). The mean actual error along with the mean initial condition error (first r.h.s term of Eq. (\ref{errore-medio-attr})) and the total error, model plus initial condition, are displayed for the configurations $C_I$ and $C_{II}$. The error fields are shown as a function of the model gridpoints at times $t=0$, $12$ and $24$ hours. The results indicate that the approximate linear evolution, Eq. (\ref{errore-medio-attr}), is able to describe the actual mean error evolution within a good level of accuracy up to $t=12$ hours, with a discrepancy between the actual error and its estimate being of the order of $10\%$ of the error amplitude. At $t=24$ hours, the linear approximation becomes less accurate.
Note that the prediction error is dominated by the model error.
Similar results (not shown) are obtained when choosing other parameter values listed in Tab. 1. 

\begin{figure*}
\centering
\includegraphics{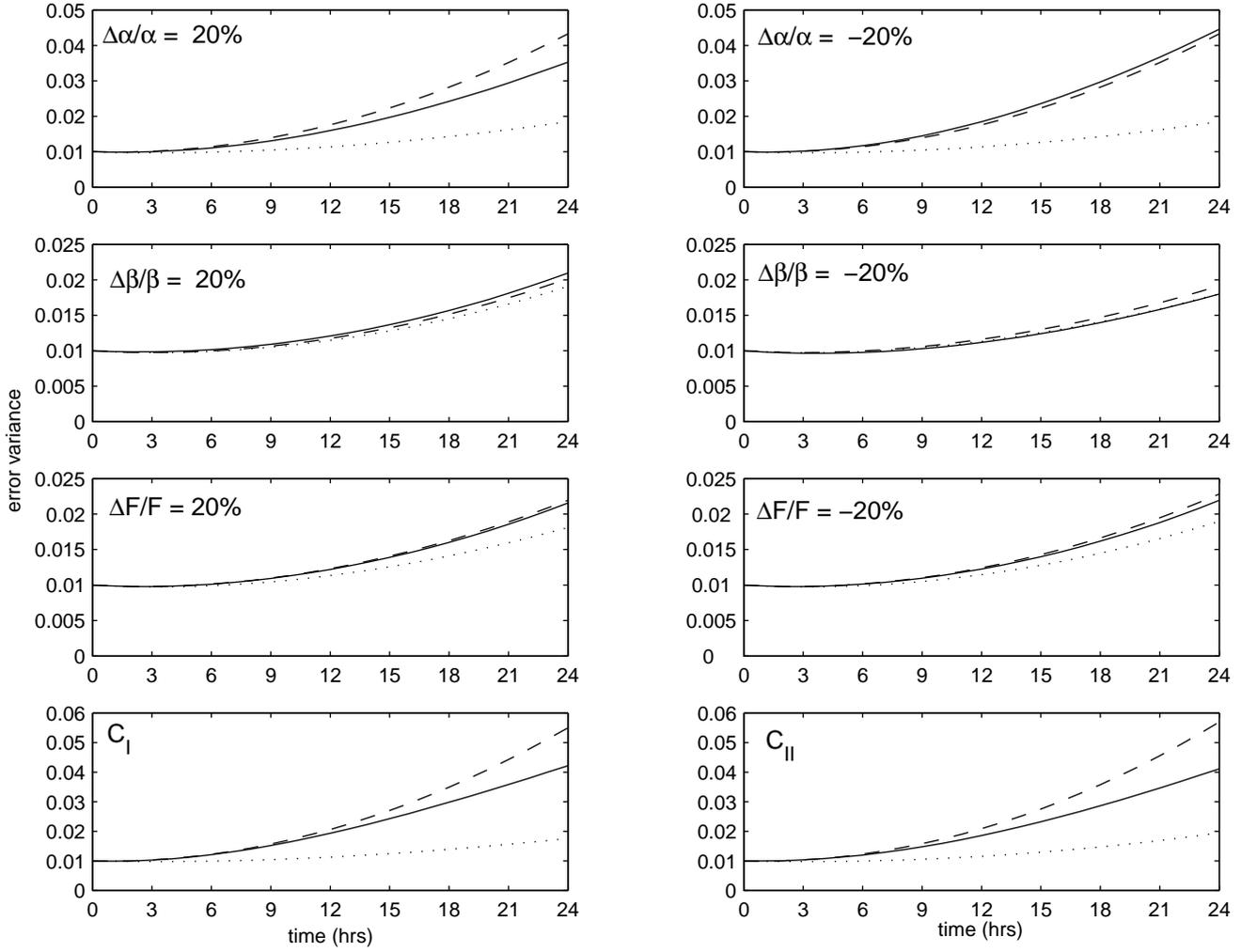}
\caption{\label{FIG2} Mean square error (trace of the covariance matrix) evolution for different model errors, indicated in each panels. Actual error (continuous line); initial condition error - linear dynamics $<({\bf M}_{t,t_0}\delta{\bf x}_0)({\bf M}_{t,t_0}\delta{\bf x}_0)^T>$ (dotted line); total error approximated by Eq. (\ref{dyn-cov-approx-q-nocorr-attr}) (dashed line). All values are normalized with the true system climate variance.}
\end{figure*}

Figure \ref{FIG2} shows the actual state estimation error variance (the trace of the covariance matrix), along with the initial condition error variance (the trace of ${\bf P}^{ic}$) and the short time error variance evolution, modeled by Eq. (\ref{dyn-cov-approx-q-nocorr-attr}). The eight plots refer to experiments with different parametric errors, according to the values given in Tab. 1. The figure indicates that in most of the cases Eq. (\ref{dyn-cov-approx-q-nocorr-attr}) provides a more accurate description of the actual quadratic error than the initial condition term alone. The estimation error variance due to model parameter misspecification grows quadratically for short times, up to a time proportional to the inverse of the most negative Lyapunov exponent (see Tab. 1), in agreement with the theory (Nicolis, 2003). For $\Delta\beta/\beta=20\%$ the addition of the quadratic model error evolution does not lead to significant difference while for $\Delta\beta/\beta=-20\%$ it slightly deteriorates the prediction. Note anyhow that for $\Delta\beta/\beta=\pm 20\%$ the impact of the model error is very small during the $24$ hours prediction. Furthermore results of Fig. \ref{FIG2} suggest that the additional terms, accounting for the correlation between model and initial condition error and that are not included in the estimate dedscribed in the figure, do not contribute much to the error dynamics in the situations considered. 

\begin{figure*}
\centering
\includegraphics{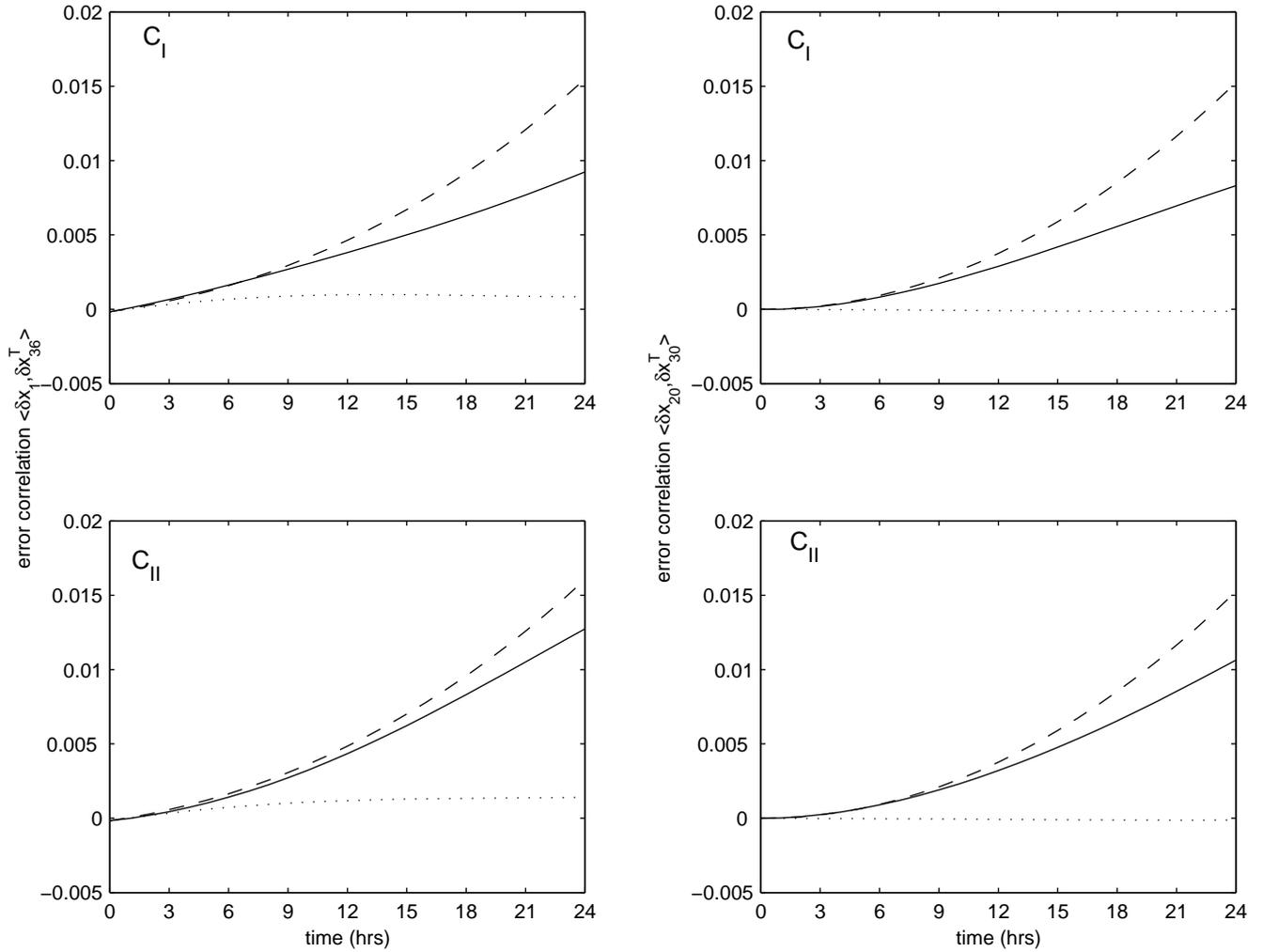}
\caption{\label{FIG3}  Error covariance evolution between points $1$ and $36$ (left panels) and $20$ and $30$ (right panels). Actual error (continuous lines); initial condition error - linear dynamics ${\bf M}_{t,t_0}<(\delta{\bf x}_0)(\delta{\bf x}_0)^T>{\bf M}_{t,t_0}^T$ (dotted lines) and the total error approximated by Eq. (\ref{P-approx}) (dashed lines). Parametric model error configuration $C_I$ (top panels) and $C_{II}$ (bottom panels). The values are normalized with the nature climate variance.} 
\end{figure*}

Figure \ref{FIG3} depicts the evolution of the error covariance between $x_1$ and $x_{36}$ and between $x_{20}$ and $x_{30}$ for the cases $C_I$ and $C_{II}$. We see that, although the accuracy is reduced with respect to the case of the error variance (Fig. \ref{FIG2}), Eq. (\ref{dyn-cov-approx-q-nocorr-attr}) appears still able to provide an accurate approximation of the actual error dynamics.

Some recent advanced data assimilation techniques are designed to track and control the instabilities which grow along a trajectory solution of a data assimilation cycle and try to optimally use the observations available to reduce the portion of the actual error which projects on the unstable direction of the system (Carrassi et al., 2007). The basic paradigm is that a relevant portion of the actual estimation error evolve according to the system's unstable subspace dynamics so that reducing the error in this subspace maximizes the overall effect of the assimilation. As a result, the analysis error is mainly confined in the complement of the subspace where the analysis increment is confined. 

In all the experiments described so far, we have used a random distribution of initial condition errors. In relation to the aforementioned advanced data assimilation algorithms, an additional relevant question concerns the accuracy of the proposed short time approximate equations, when the initial condition errors (i.e. the analysis errors) are confined to the unstable/stable subspace of the model's solution. To this aim experiments similar to the ones described above have been performed, except that the sampling of the initial condition error is made after an initial transient of $2$ years necessary to compute the Lyapunov exponents of the system. The latter, as well as the Lyapunov vectors, are estimated using the standard Gram-Schmidt procedure (Benettin and Galgani, 1980). 
The initial condition errors are then sampled with the constraint of being either aligned to the first Lyapunov vector or orthogonal to the local unstable subspace (in practice being aligned with the direction associated to the most negative Lyapunov exponent). The results (not shown) for the parametric error configurations $C_I$ and $C_{II}$, indicate that although the initial condition error evolution is clearly conditioned by constraining the initial condition error to the stable/unstable subspace, the effect on the accuracy and duration of the approximate model error quadratic regime is negligible.   

In conclusion, the numerical experiments indicate that, in the short time (within 24 hours), Eqs. (\ref{errore-medio-attr}) and (\ref{dyn-cov-approx-q-nocorr-attr}) may be used to model the mean and covariance error evolution within the assimilation interval in a sequential assimilation scheme. 

\subsection{EKF in the presence of parametric model error}

OSSEs with the EKF will next be performed in the context of the Lorenz 36-variables (Lorenz, 1996). An homogeneous network of 18 observations is used at which model variables are observed $(x_i$, $i=1,3,...,35)$. The simulated observations are generated by sampling the reference true trajectory with an unbiased Gaussian random measurement error of mean zero and variance $\sigma_o^2$. The observation error variance is $\sigma_o^2=2.5\%$ of the system's climate variance and three assimilation intervals, $\tau$, are considered: $12$, $6$ and $3$ hours. The model is represented by Eq. (\ref{model-expr}) with the parameters given in Tab. 1. All the experiments are performed for $6$ years; the first year of simulated time is considered to as a spin-up period so that all results and statistics refer to the last 5 years.  

First, we report the performance of the filter, in terms of the accuracy of its state estimate, when a perfect model is used and consequently no model error treatment is employed in the EKF ({\it i.e.} ${\bf P}^{m}=0$ in Eq. (\ref{Pf_EKF})). 
It is known that because of the nonlinearities, even in a perfect model scenario, the EKF solution may drift away from the true trajectory (Miller et al., 1994). A few empirical solutions have been proposed in previous studies with the EKF in order to reduce or avoid the filter divergence; namely the multiplicative forecast error variance inflation (Anderson, 2001) or the addition of random perturbations to the diagonal of the analysis error covariance matrix to enhance its explained variance (Corazza et al., 2003; Yang et al., 2006). In our experiments with the EKF we have opted for the latter solution; random perturbations $\delta$ are added to the diagonal of the analysis error covariance matrix after the analysis update, Eq. (\ref{Pa_EKF}): $\delta=\xi\alpha\sigma_o^2$, with $0<\xi\leq1$ a random number extracted from a normal distribution and $0<\alpha\leq1$ a tunable scalar coefficient. The latter, optimized by minimizing the average EKF analysis error over 5 years, is equal to $0.2$.

When a perfect model is used, the overall performance of the filter is good. The time mean analysis error variance, over the 5 years, is equal to $0.89\%$, $0.76\%$ and $0.66\%$ of the system climate variance, with assimilation intervals equal to $12$, $6$ and $3$ hours respectively.

\begin{figure*}
\centering
\includegraphics{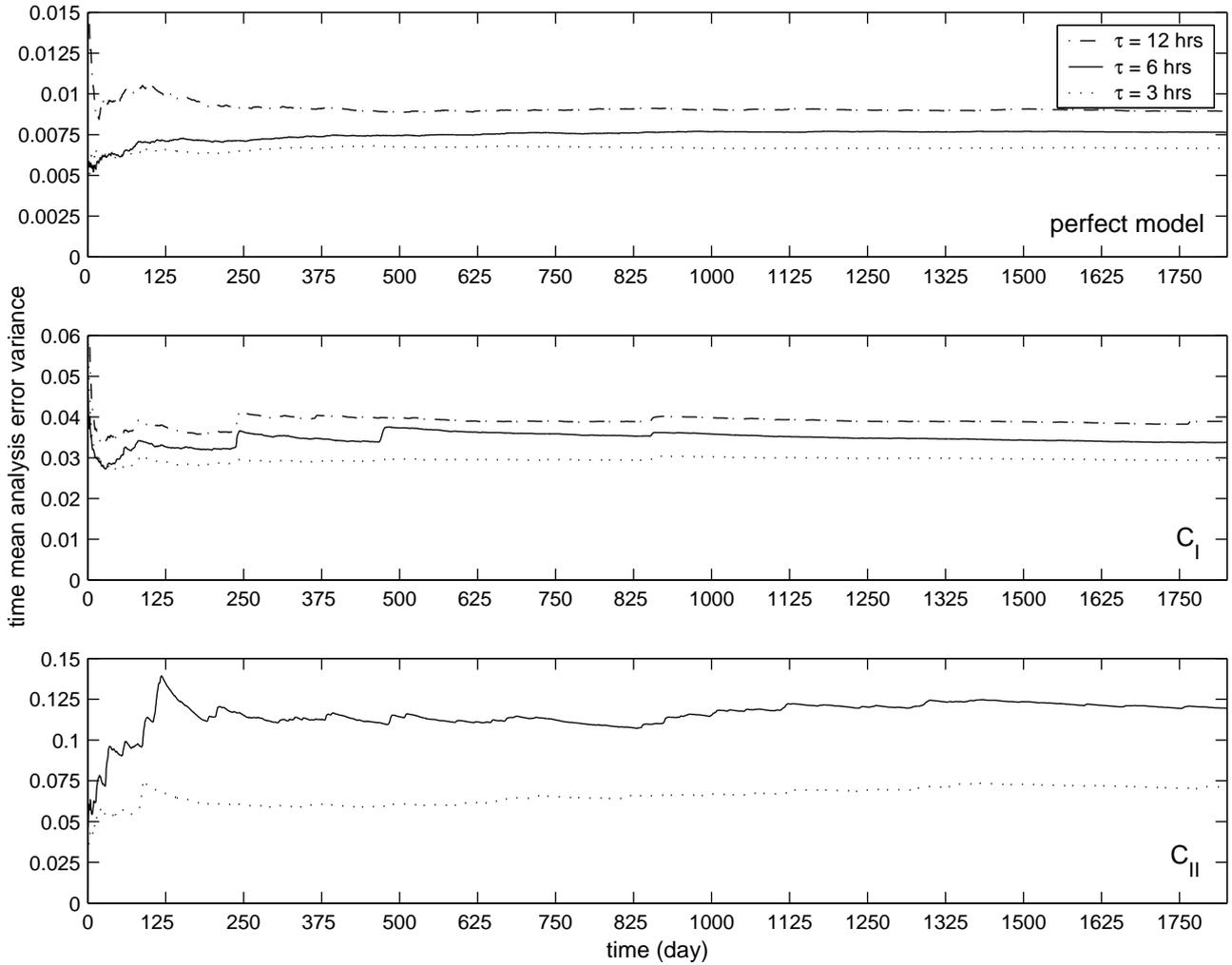}
\caption{\label{FIG4}  Time running mean of the EKF analysis error variance. The EKF is employed in the perfect model scenario applied on an imperfect model. Parametric model error configurations perfect (top panel), $C_I$ (mid panel), $C_{II}$ (bottom panel). Assimilation intervals, $\tau =$ 12 hours (dash-dotted lines), 6 hours (continuous lines) and 3 hours (dashed lines). Values are normalized with the nature climate variance.}
\end{figure*}

We now turn to introduce parametric errors in the model. As a first step, the filter performance without model error treatment is analyzed. Figure \ref{FIG4} shows the time running mean of the analysis error variance as a function of time, for experiments in which the model is in configuration $C_I$ and $C_{II}$. The perfect model case is also displayed for reference.     
The impact of model error is dramatic: in the configuration $C_I$ the average analysis error is about four times the corresponding error when a perfect model is used, while it is more than one order of magnitude in the configuration $C_{II}$; in this latter case filter divergence occurs when $\tau=12$ hours. In all the cases, the average analysis error is larger than the observation error variance even when the shortest assimilation interval, $\tau = 3$ hours, is used.  
Besides the increased mean error level, the curves show abrupt jumps (in particular when $\tau=12$ or $6$ hours) indicating that, in the presence of model error, the filter undergoes large error fluctuations during the analysis cycle. Large error spikes may be also due to changes in the system regime which are not tracked on by the filter solution.
Similar experiments, with the EKF assuming a perfect model scenario, have been performed for all the model parameters configurations given in Tab. 1 and the corresponding time average analysis error are summarized in Tab. 2.
As expected, the presence of the model error deteriorates systematically the performance of the filter. 

Before implementing the EKF with the model error covariance matrix estimated through Eq. (\ref{PMapprox}) or (\ref{PMWNapprox}), the question on the accuracy of the linearized model error evolution within the analysis intervals is addressed in a more idealized experimental setting. We assume to know the model error at each analysis time, $\delta\mu^{a} = \delta\mu_{0}$, so that the model error at forecast time is estimated according to Eq. (\ref{dyn-err-der_Ib}), that is:
\begin{equation}
\label{mod_TD}
\delta{\bf x}^m_{k+1}\approx\delta{\bf\mu}^a_k(t_{k+1}-t_k)=\delta{\bf\mu}^a_k\tau
\end{equation}
where $t_k$ indicate an arbitrary analysis time along the assimilation cycle; $\delta{\bf\mu}^a_k$ is evaluated through Eq. (\ref{model_error}) by using the analysis state ${\bf x}^a$. This estimate of the model error is then removed from the background (forecast) field, before the EKF analysis update as for the bias removal, Eq. (\ref{back_rm}), except that now a time-dependent estimate of the model error is used and ${\bf P}^m=0$ in Eq. (\ref{Pf_EKF}). This simplifies the interpretation of the results since, by avoiding the uncertainty associated to the estimate of the model error covariance matrix, the main source of errors is the limited accuracy of the approximate model error dynamics Eq. (\ref{mod_TD}). The EKF employing this time-dependent model error removal has been tested with the model in the configurations $C_I$ and $C_{II}$. In both cases, the time average analysis error is reduced below the observation error variance. The time average analysis errors, in the configuration $C_I$ are $2.34\%$, $1.02\%$ and $0.71\%$ with assimilation interval as long as $12$, $6$ and $3$ hours respectively, while in the configuration $C_{II}$ they are $1.47\%$ and $0.74\%$ with $\tau=6$ and $3$ hours respectively. Note that in the case $C_{II}$ and $\tau=12$ hours the filter diverges suggesting that the short time approximation of the model error evolution is not sufficiently reliable on this time scale. The overall results are encouraging: by employing this idealized time-dependent model error removal the EKF analysis errors attain values comparable to those obtained with a perfect model (see the first line of Tab. 2).

We now turn to investigate the performance of the extensions of the EKF proposed in Sect. 3, in which the forecast error covariance matrix, ${\bf P}^f$, used in the EKF analysis update, includes also a representation of the model error covariance. 
The latter is estimated by Eq. (\ref{PMWNapprox}) for the white noise assumption and by Eq. (\ref{PMapprox}) for the deterministic case. The ${\bf P}^m$ matrix is kept constant along all the EKF analysis cycle. Furthermore in both cases, the bias, estimated through Eq. (\ref{bias_es}), is removed from the background field before the latter is used in the analysis update. The model error mean, $<\delta{\bf\mu}>$, and covariance, $<(\delta{\bf\mu}-<\delta{\bf\mu}>)(\delta{\bf\mu}-<\delta{\bf\mu}>)^T>$, to be used in Eqs. (\ref{bias_es}), (\ref{PMapprox}) and (\ref{PMWNapprox}) are evaluated, for each of the parameters values described in Tab. 1, by accumulating statistics over a sample of $10^5$ initial conditions on the system's attractor.

\begin{figure*}
\centering
\includegraphics{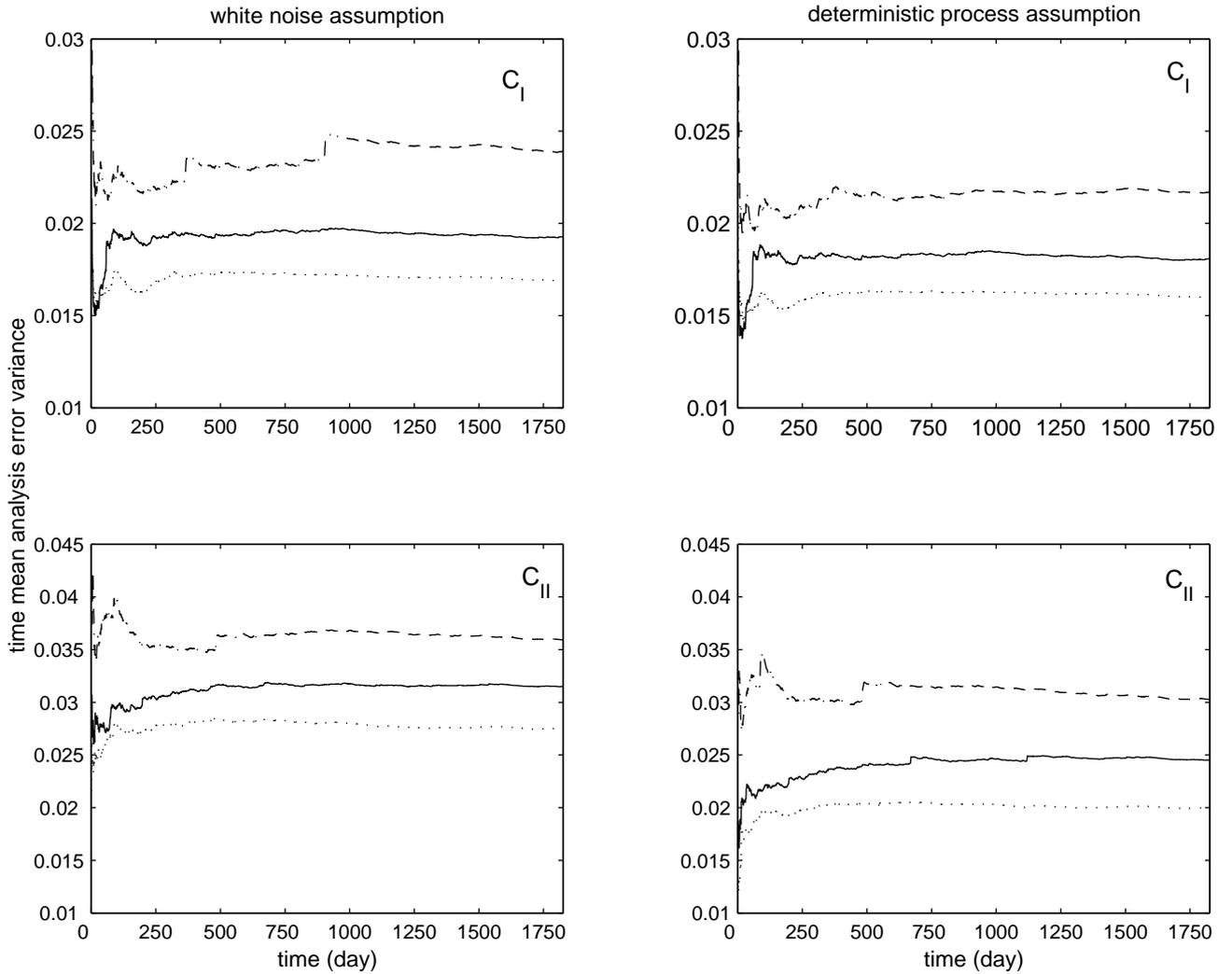}
\caption{\label{FIG5} Time running mean EKF analysis error variance for which model error is considered as a white noise, Eq. (\ref{PMWNapprox}) (left column) and a deterministic process, Eq. (\ref{PMapprox}) (right column). Parametric model error configurations: $C_I$ (top panels) and $C_{II}$ (bottom panels). The assimilation interval is fixed to $\tau =$ 12 hours (dash-dotted lines), 6 hours (continuous lines) and 3 hours (dashed lines). The values are normalized with the nature climate variance.} 
\end{figure*}

Figure \ref{FIG5} shows the running mean EKF analysis error as a function of time, in the white noise and deterministic process assumptions, with assimilation intervals $\tau=12$, $6$ and $3$ hours. The plots refer to the model parametric configurations $C_I$ and $C_{II}$. In both situations, the best performance is obtained when the model error covariance matrix is built on the basis of a quadratic error evolution law between assimilation times. In the case of the more unstable configuration $C_{II}$, the improvement is still more pronounced and, for instance, when the model error is treated as a white noise, a $3$ hours assimilation interval is needed to reach mean error values smaller than those obtained in the deterministic case with an assimilation interval of $12$ hours. It is furthermore relevant to note that most of the analysis error jumps apparent in the running mean, are significantly reduced by employing the deterministic model error description.  

Other analysis cycle experiments have been performed for all the parametric configurations given in Tab. 1 and the results are reported in Tab. 3 and in Tab. 4 for the white noise and deterministic case respectively. For almost all the model parameters considered, best results are obtained when the model error is treated as a deterministic process: the only exception is the case in which the actual dissipation parameter is underestimated ($\frac{\Delta\bf\beta}{\bf\beta}=20\%$) and the assimilation interval equal to $\tau=12$ hours. We have seen in Section 4.1 that this is one of the case in which the model error has only a minor impact on the prediction error. In this circumnstance the accuracy of the filter solution is mainly related to the accuracy of the forecast error description based on the linear propagation of the estimated analysis error. We argue that the addition of a larger model error covariance matrix (such as that corresponding to the white noise assumption: ${\bf P}^m_{wn}={\bf Q}\tau > {\bf P}^m_{dp}={\bf Q}\tau^2$) helps to better preventing possible underestimation of the actual forecast error covariance occuring in the presence of strong nonlinearities.

\begin{figure*}
\centering
\includegraphics{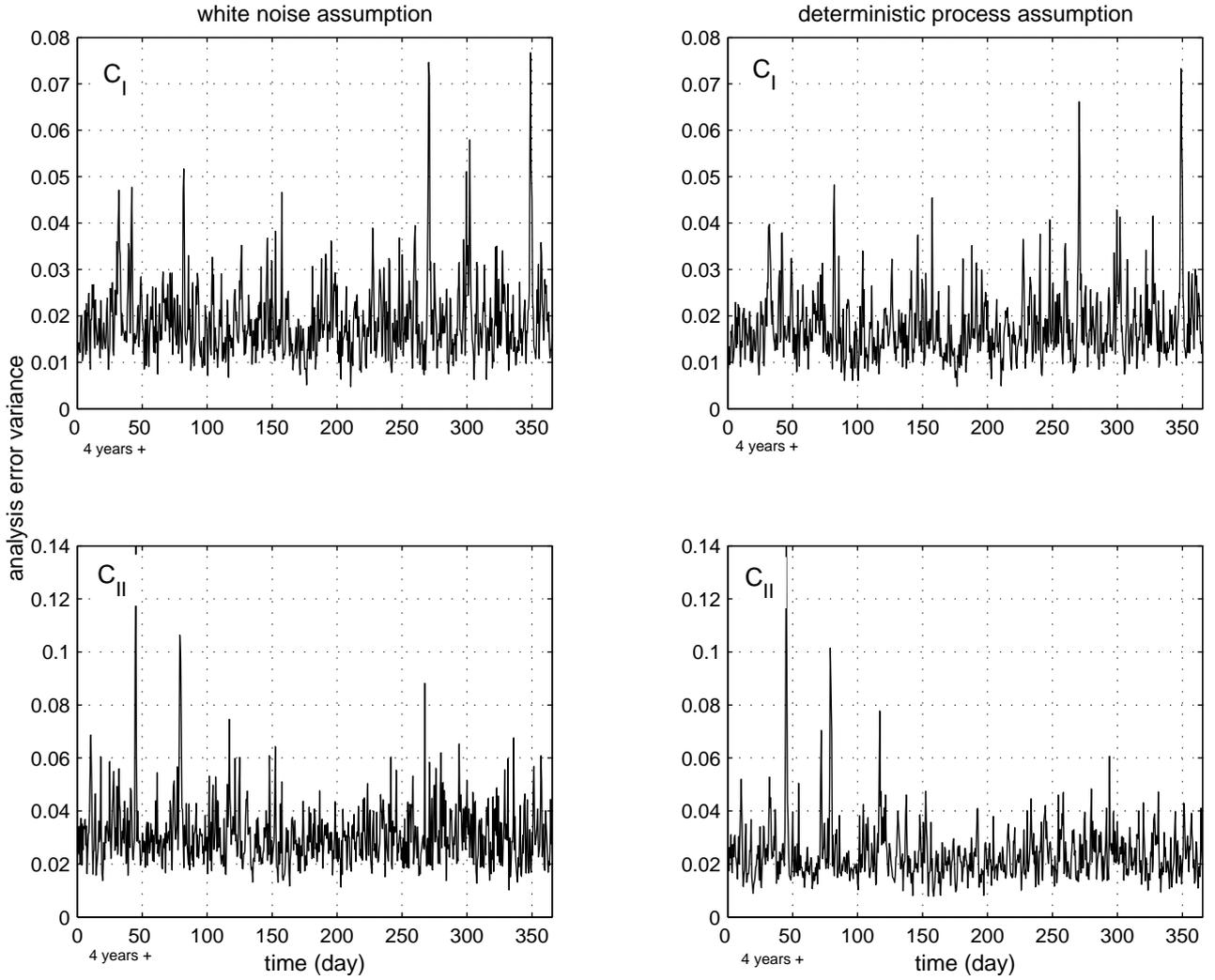}
\caption{\label{FIG6}  EKF analysis error variance as a function of time for the experiment with $\tau=$6 hours, during the last year of the simulation. Model error treatment: white noise hypothesis, Eq. (\ref{PMWNapprox}) (left column), deterministic process, Eq. (\ref{PMapprox}) (right column). Parametric model error configurations:  $C_I$ (top panels) and $C_{II}$ (bottom panels). The values are normalized with the nature climate variance.} 
\end{figure*}

Figure \ref{FIG6} illustrates the EKF analysis error variance as a function of time for the experiments of Fig. \ref{FIG5} with $\tau=6$ hours; the figure displays the analysis error during the last year of simulated time. As already evident from Fig. \ref{FIG5} the average analysis error is lower in the deterministic case. From Fig. \ref{FIG6} we further see that the variability about the mean values appears slightly smaller too and, except for a few instances, error peaks are all reduced by treating model error as a deterministic process. 

\begin{figure*}
\centering
\includegraphics{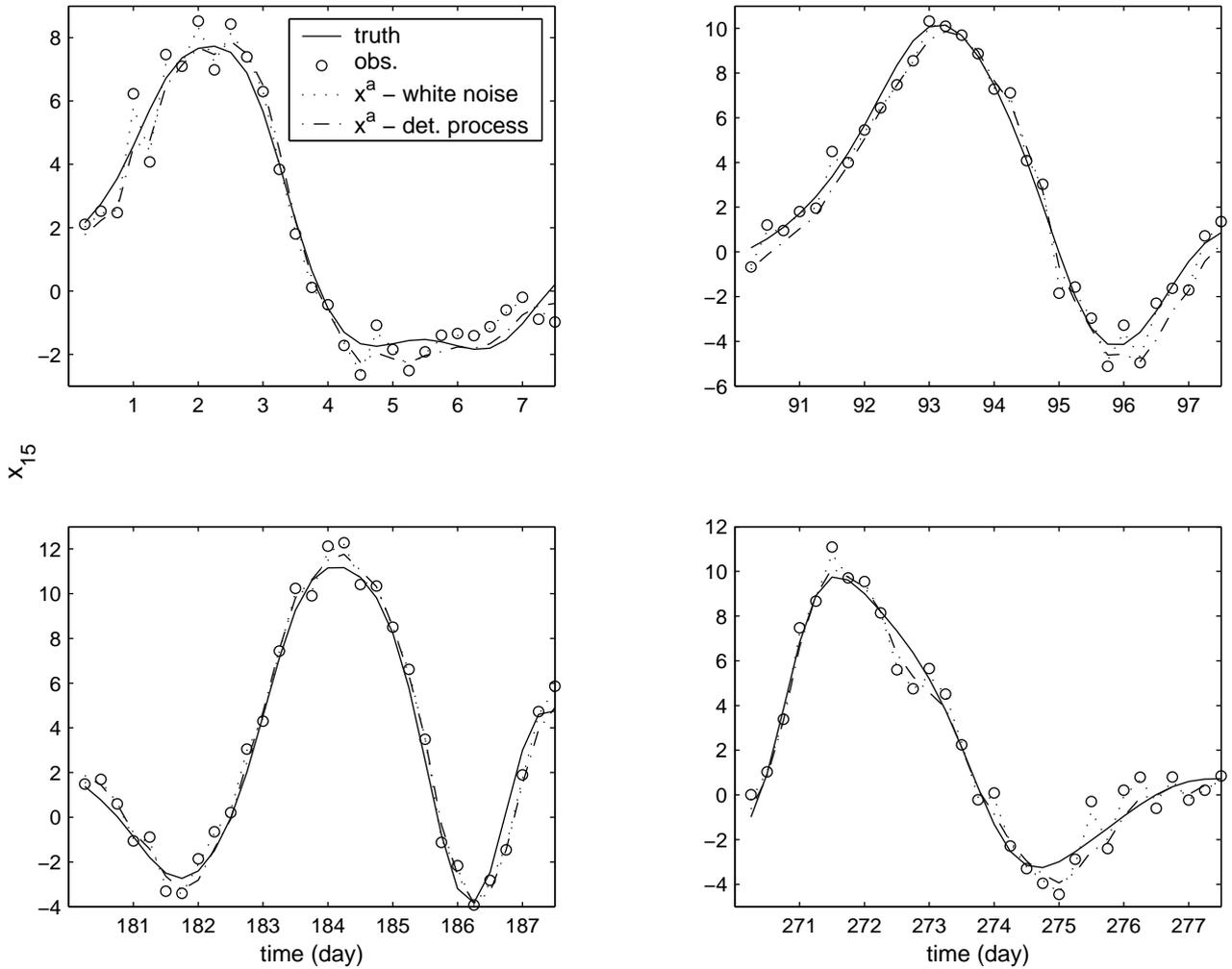}
\caption{\label{FIG7} EKF experiments with $\tau=$6 hours - $x_{15}$ as a function of time for model configuration $C_{II}$. The true system (continuous lines), observations (circles), the EKF solution with the white noise hypothesis for model error (dotted lines) and th EKF solution with the deterministic process hypothesis for model error (dash-dotted lines), are plotted. The four plots display the field values during the first week of the 1$^{th}$, 4$^{th}$, 7$^{th}$ and 10$^{th}$ month of the last year of simulated time.} 
\end{figure*}

A further illustration of the ability of the filter to track the nature evolution is given in Fig. \ref{FIG7} which shows the true value at $x=15$, the observation, taken every $6$ hours, and the two filter solutions in the case of white noise or deterministic model error. The four plots display the field values during the first week of the 1$^{th}$, 4$^{th}$, 7$^{th}$ and 10$^{th}$ month of the last year of simulated time. There are evidences of instances, particularly in correspondence with the maximum and minimum, in which the filter with the deterministic model error formulation appears to provide a smoother solution than in the white noise assumption. In some cases (for instance around day $2$ in the left top panel, or day $271$ in right bottom panel) the analysis solution in the deterministic case remains closer to the true solution in spite of the large observation error. This indicates the accurate estimation of the forecast error accomplished by modeling model error as a deterministic process. The white noise assumption, on the other hand, seems to lead to the overestimation of the actual forecast error so that the analysis gives a larger, improper, weight to the observations. 

\begin{figure*}
\centering
\includegraphics{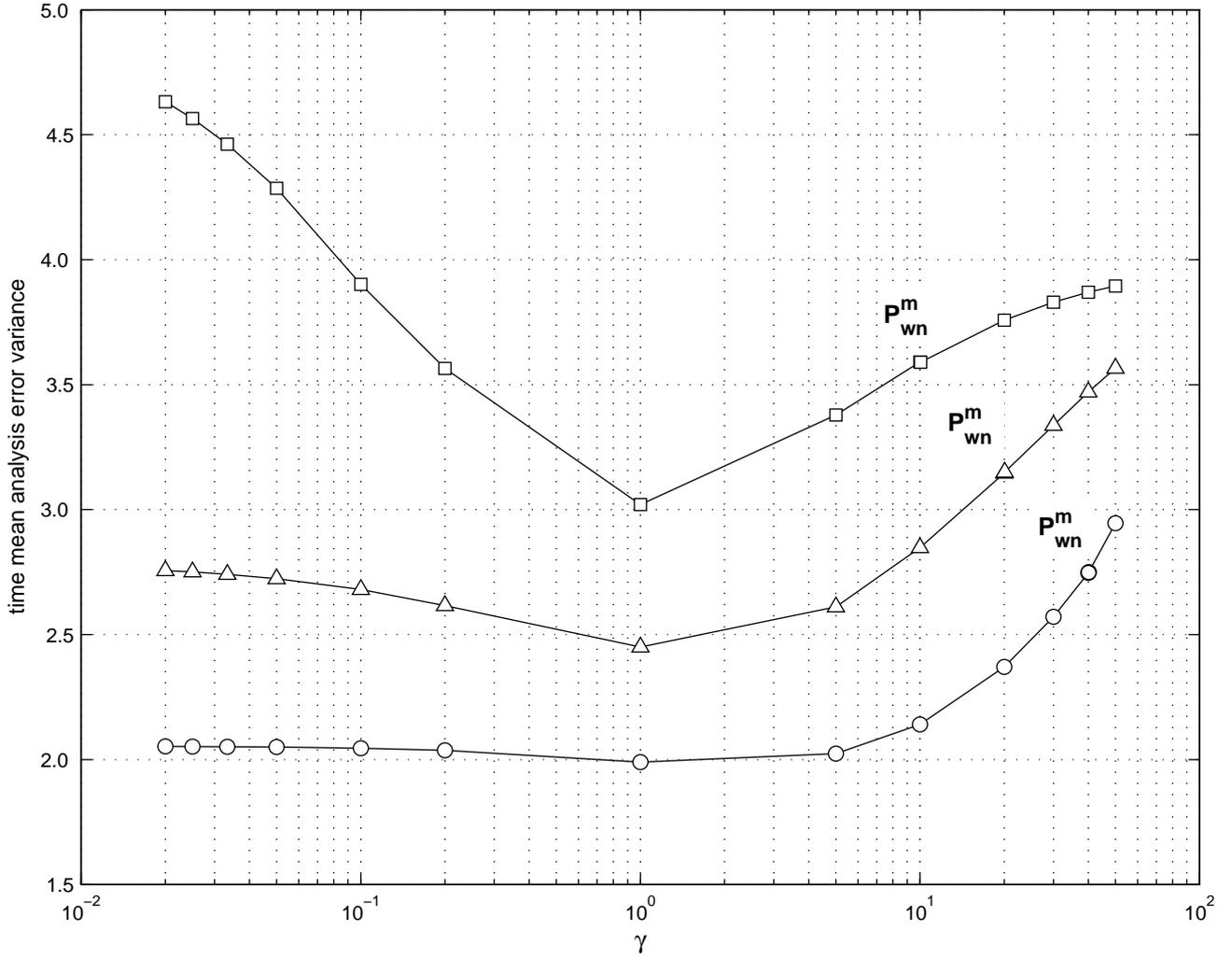}
\caption{\label{FIG8}  EKF time average analysis error variance as a function of the coefficient $\gamma$ of the model error covariance matrix, ${\bf P}^m=\gamma{\bf P}^m_{dp}$. The parametric model error configuration is $C_{II}$ and the assimilation interval is fixed to $\tau=3$ hours (circles), $\tau=6$ hours (triangles) and $\tau=12$ (squares). The values are normalized with the nature climate variance. Note that the x-axis is in logaritmic scale.} 
\end{figure*}

To further examine the quality of the deterministic approach, it is interesting to study the performance of the filter by tuning the amplitude of the model error covariance used in the EKF, that is ${\bf P}^m=\gamma{\bf P}^m_{dp}$. This allows to evaluate, on an heuristic basis and a-posteriori, how far from optimality is the given model error covariance matrix. Figure \ref{FIG8} shows the time average analysis error as a function of the scalar coefficient $\gamma$, for the three different assimilation intervals, $\tau=3$, $6$ and $12$ hours and for the configuration $C_{II}$. Remarkably, for all the assimilation intervals considered, the minimum average analysis error is attained when ${\bf P}^m={\bf P}^m_{dp}$. For reference, the values corresponding to the EKF under the white noise hypothesis are labeled in the figure to allow a direct evaluation of the improvement gained by employing a deterministic representation of the model error. Note that, once an optimal model error covariance matrix has been estimated for a given assimilation interval, the optimal amplitude of the model error covariance matrix can be estimated on the basis of the quadratic time evolution for any (short) assimilation interval.     

\begin{figure*}
\centering
\includegraphics{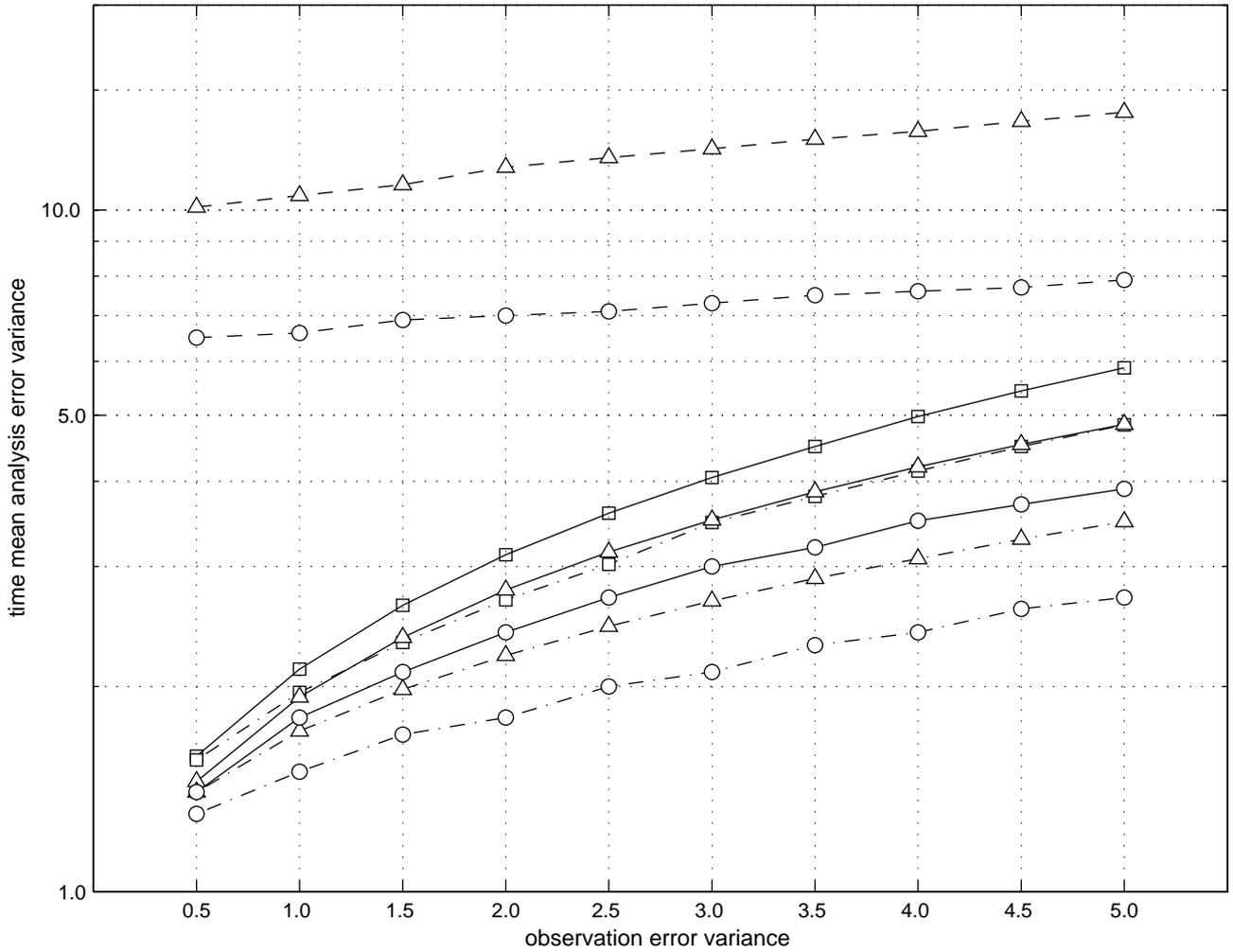}
\caption{\label{FIG9} EKF time average analysis error variance as a function of the observation error variance (expressed as a percentage of the system's climate variance) for parametric model error configuration $C_{II}$. The assimilation interval is fixed to $\tau=3$ hours (circles), $\tau=6$ hours (triangles) and $\tau=12$ (squares). The different lines refer to the perfect assumption (dashed lines), the white noise assumption (continuous lines) and the deterministic process assumption (dash-dotted lines). The values are normalized with the nature climate variance. Note that the y-axis is in logaritmic scale.} 
\end{figure*}

Our analysis has so far focused on examining the performance of the filter with different analysis intervals. This was mainly motivated by the interest in exploring the range of validity of the short time approximation on which the model error formulation relies. The following experiments, on the other hand, are aimed at a sensitivity analysis. The robustness of the EKF is examined by modifying the properties of the observational network, more specifically the observation error amplitude and the number and location of the observations. Figure \ref{FIG9} is obtained by using the same homogeneous distribution of 18 observations used so far, but the observation error variance is varied from $0.5\%$ to $5\%$ of the system's climate variance. The model is in the configuration $C_{II}$. Three types of EKF analysis cycles are compared: $(i)$ a perfect model scenario, $(ii)$ a white noise model error and $(iii)$ a deterministic model error. The figure shows the time average analysis error over the last 5 years of the experiments as a function of the observation error variance, with an assimilation interval $\tau=12$, $6$ and $3$ hours. For all the observation error values and assimilation intervals considered, best performances are obtained when model error is treated as a deterministic process. In the case the model error is not accounted for, and the assimilation interval is $12$ hours, the filter diverges for all the values of observation errors. Note that by reducing the observation error, the relative difference between the random or deterministic approaches reduces too. This is reasonable since, when observation error is very small, the resulting analysis is very accurate and the forecast error covariance matrix ${\bf P}^f$ is dominated by the model error component ${\bf P}^m$. In this condition, ${\bf P}^f \approx {\bf P}^m >> {\bf R}$ and the analysis does not depend on the amplitude of the forecast error covariance matrix, as if the observations were perfect, (see Eq. \ref{analysis-update}).  

\begin{figure*}
\centering
\includegraphics{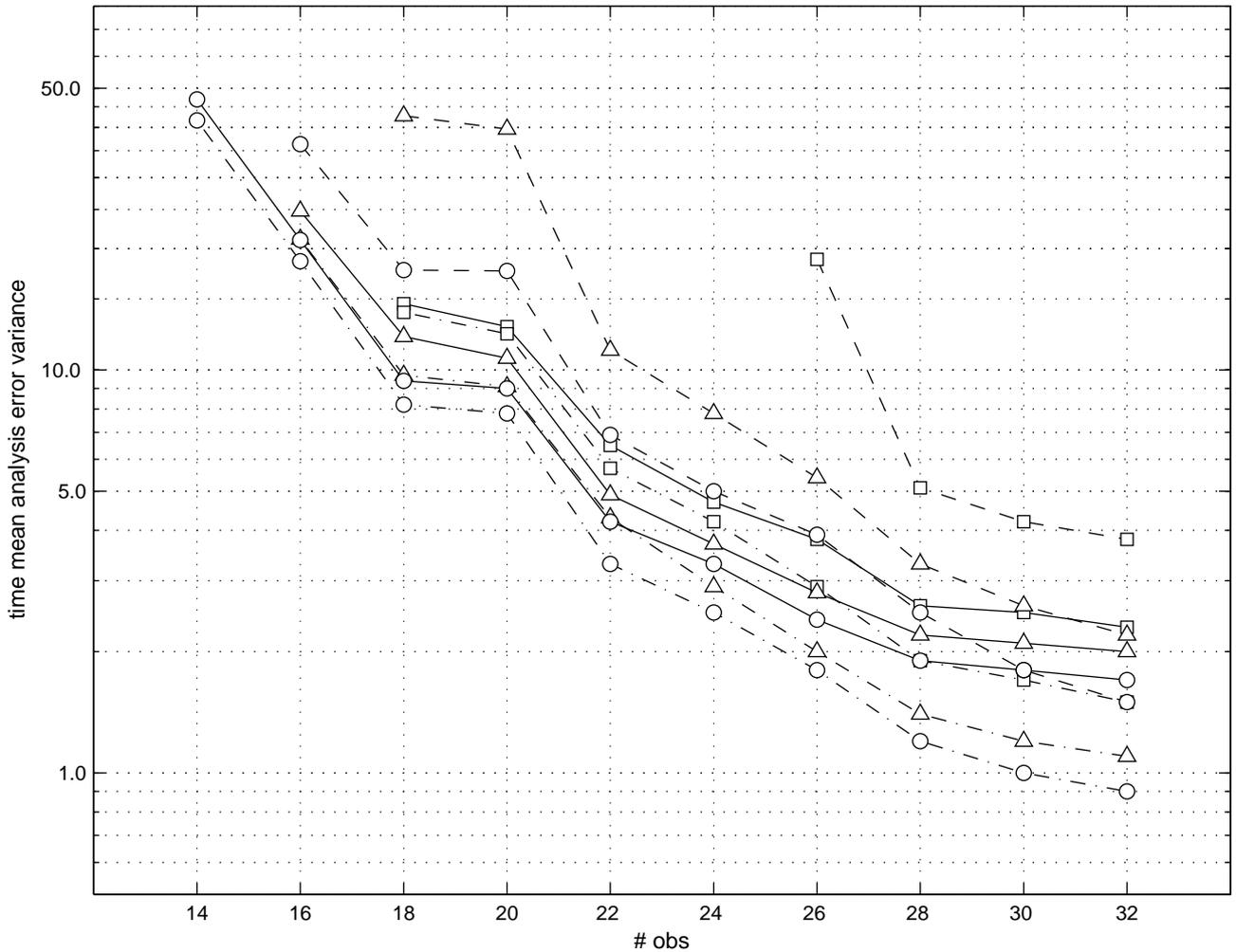}
\caption{\label{FIG10} EKF time average analysis error variance as a function of the number of observations for the parametric model error configuration $C_{II}$. The assimilation interval is fixed to $\tau=3$ hours (circles), $\tau=6$ hours (triangles) and $\tau=12$ (squares). The lines refer to the perfect assumption (dashed lines), the white noise assumption (continuous lines) and the deterministic process assumption (dash-dotted lines). The values are normalized with the nature climate variance. Note that the y-axis is in logaritmic scale.} 
\end{figure*}

In Fig. \ref{FIG10}, the time average analysis error is plotted as a function of the number of observations. In these experiments, a random distribution of observations is deployed; their number ranges from 14 to 32 and for each of these values, 10 random distributions of observations are generated, and analysis cycles with the EKF are performed. The results are averaged in space and time as well as over the 10 observation distributions. Observation error variance is fixed to $2.5\%$ of the system climate variance. The deterministic model error treatment is still the best for all the cases considered. It is interesting to note the increasing advantage gained with the deterministic assumption when the number of observations is increased too: it seems that, as long as the observing network is refined, the average estimation error is reduced and the linear hypothesis on which the matrix ${\bf P}^m_{dp}$ is built becomes more accurate. 

\section{Summary and Conclusion}

Data assimilation constitutes nowadays a central part of the operational weather forecasting system. The purpose is to get an estimate of the state of the atmosphere as close as possible to reality and compatible with the simulating model at hand. To this end accurate information on the observations and on the model are necessary. A central problem is the discrepancy between the model and the true dynamics, the model error, for which simplistic assumptions are usually made such as, for instance, that of a white noise process.

In the present work a different approach has been adopted, based on recent advances made on the dynamics of model errors. Our analysis does not make use of any a-priori assumption on the model error dynamics except its deterministic character implying the existence of a short term universal behavior of model errors deduced in Nicolis (2003; 2004). The analytical development of evolution equations for the mean and covariance error, in the presence of both initial condition and model uncertainties, has shown the way this deterministic approach affects the structure of the classical extended Kalman filter equations. 

The theoretical analysis is fully supported by numerical experiments in the context of a low order atmospheric model giving rise to chaotic behavior (Lorenz, 1996). Model errors are explicitly introduced in this system by perturbing the parameters with respect to some reference true values. It is shown that in the short time (less than 24 hours) the estimation error is accurately approximated by an evolution law in which the model error, treated as a deterministic process, is expanded in a Taylor series in time up to the first nontrivial order. The correlation between model and initial condition error has only a minor impact on the short time evolution of the error covariance matrix. Furthermore the possibility of using the deterministic approach to account for the model error dynamics in the extended Kalman filter has been explored and compared to the classical white noise approach. To this end, observations system simulation experiments with the extended Kalman filter have been performed in the Lorenz model (Lorenz, 1996). The numerical analysis allowed us to assert that there is a clear indication of a substantial improvement of the filter accuracy (in terms of analysis error variance) when model error is assumed to act as a deterministic process. 
The filter performance has been further examined by varying either the assimilation intervals or the observational network properties ({\it i.e.} the observation number and error variance). As long as the assimilation interval is short enough the filter employing the quadratic model error evolution law (deterministic process assumption) outperforms systematically the white noise case. 

The existence of this universal, short time, quadratic law for the evolution of model error covariance might turn to be useful in all the situations in which access to some statistical information on the parametric model error is available and can then be used to estimate the model error covariance matrix. At the same time, if, for a given assimilation interval, an optimal model error covariance matrix is at hand, the optimal matrix at different assimilation intervals can be evaluated straightforwardly on the basis of the quadratic law. 

It would be interesting to apply the deterministic approach outlined in this work to models of increasing complexity and for parametric errors which have a direct physical interpretation. Furthermore, a number of more fundamental questions remain to be addressed concerning the way to account for other types of model errors. In the present study, the focus was placed on parametric errors assuming that the model and the reference system span the same phase space. In a realistic setting, such as for instance numerical weather prediction, errors arise also from the effect of the unresolved scales as well as from the lack of the description of some relevant phenomena. These problems will be addressed in future work.

\section*{Acknowledgement}
This work was supported by the Belgian Federal Science Policy Program under contract MO/34/017.

\newpage


\newpage

\begin{table}[hc]
\label{tab1}
\begin{tabular}{||c|c|c|c|c|c|c|c||}
\hline\hline
$\bm\alpha$ & $\bm\beta$ & ${\bf F}$ & $\frac{\Delta{\bm\lambda}}{{\bm\lambda}}$ & ${\bf\sigma_{max}}$ & ${\bf\sigma_{min}}$ & ${\bf N_{\sigma^+}}$ & ${\bf KY-dim}$ \\ \hline \hline
\hline 1 & 1 & 8 & $\frac{\Delta{\bm\alpha}}{\bm\alpha}=0$, $\frac{\Delta{\bm\beta}}{\bm\beta}=0$, $\frac{\Delta{\bf F}}{{\bf F}}=0$  & 0.33 & -0.97 & 11 & 24.35 \\
\hline 1.2 & 1 & 8 & $\frac{\Delta{\bm\alpha}}{\bm\alpha}=-20\%$, $\frac{\Delta{\bm\beta}}{\bm\beta}=0$, $\frac{\Delta{\bf F}}{{\bf F}}=0$  & 0.32 & -0.92 & 12 & 24.29 \\
\hline 0.8 & 1 & 8 & $\frac{\Delta{\bm\alpha}}{\bm\alpha}=20\%$, $\frac{\Delta{\bm\beta}}{\bm\beta}=0$, $\frac{\Delta{\bf F}}{{\bf F}}=0$  & 0.36 & -1.03 & 12 & 25.36  \\
\hline 1 & 1.2 & 8 & $\frac{\Delta{\bm\alpha}}{\bm\alpha}=0$, $\frac{\Delta{\bm\beta}}{\bm\beta}=-20\%$, $\frac{\Delta{\bf F}}{{\bf F}}=0$ & 0.23 & -0.89 & 11 & 22.40  \\
\hline 1 & 0.8 & 8 & $\frac{\Delta{\bm\alpha}}{\bm\alpha}=0$, $\frac{\Delta{\bm\beta}}{\bm\beta}=20\%$, $\frac{\Delta{\bf F}}{{\bf F}}=0$  & 0.43 & -1.05 & 11 & 25.49  \\
\hline 1 & 1 & 9.6 & $\frac{\Delta{\bm\alpha}}{\bm\alpha}=0$, $\frac{\Delta{\bm\beta}}{\bm\beta}=0$, $\frac{\Delta{\bf F}}{{\bf F}}=-20\%$  & 0.43 & -1.09 & 12 & 26.19  \\
\hline 1 & 1 & 6.4 & $\frac{\Delta{\bm\alpha}}{\bm\alpha}=0$, $\frac{\Delta{\bm\beta}}{\bm\beta}=0$, $\frac{\Delta{\bf F}}{{\bf F}}=20\%$  & 0.22 & -0.84 & 10 & 21.35  \\
\hline 1.2 & 1.2 & 6.4 & $C_{I}: \frac{\Delta{\bm\alpha}}{\bm\alpha}=-20\%$, $\frac{\Delta{\bm\beta}}{\bm\beta}=-20\%$, $\frac{\Delta{\bf F}}{{\bf F}}=20\%$  & 0.14 & -0.76 & 8 & 17.96  \\
\hline 0.8 & 0.8 & 9.6 & $C_{II}: \frac{\Delta{\bm\alpha}}{\bm\alpha}=20$, $\frac{\Delta{\bm\beta}}{\bm\beta}=20\%$, $\frac{\Delta{\bf F}}{{\bf F}}=-20\%$  & 0.63 & -1.27 & 13 & 28.08  \\
\hline
\end{tabular}
\caption{Summary of the model parameters values used in the experiments. The parametric model error is expressed as a relative percentage of the correct, "true", parameters, ${\bm\lambda}=(\bm\alpha,\bm\beta,{\bf F})=(1,1,8)$. The leading and smallest Lyapunov exponents $\sigma_{max}$ and  $\sigma_{min}$ are expressed in (day)$^{-1}$. ${\bf N_{\sigma^+}}$: number of positive Lyapunov exponents; ${\bf KY-dim}$ Kaplan-Yorke dimension. Results are based on a 200 years long integration after a 10 years long spin-up period.} 
\end{table}

\begin{table}[hc]
\label{tab2}
\begin{tabular}{||c|c|c|c|c|c||}
\hline\hline
$\frac{\Delta{\bm\lambda}}{{\bm\lambda}}$  & $\tau=12$ hrs & $\tau=6$ hrs & $\tau=3$ hrs  \\ \hline \hline
\hline $\frac{\Delta{\bm\alpha}}{\bm\alpha}=0$, $\frac{\Delta{\bm\beta}}{\bm\beta}=0$, $\frac{\Delta{\bf F}}{{\bf F}}=0$ & 0.89\% & 0.76\% & 0.66\%   \\
\hline $\frac{\Delta{\bm\alpha}}{\bm\alpha}=-20\%$, $\frac{\Delta{\bm\beta}}{\bm\beta}=0$, $\frac{\Delta{\bf F}}{{\bf F}}=0$ & 2.55\% & 1.95\% & 1.71\%   \\
\hline $\frac{\Delta{\bm\alpha}}{\bm\alpha}=20\%$, $\frac{\Delta{\bm\beta}}{\bm\beta}=0$, $\frac{\Delta{\bf F}}{{\bf F}}=0$ & 4.16\% & 2.85\% & 2.24\%   \\
\hline $\frac{\Delta{\bm\alpha}}{\bm\alpha}=0$, $\frac{\Delta{\bm\beta}}{\bm\beta}=-20\%$, $\frac{\Delta{\bf F}}{{\bf F}}=0$ & 1.09\% & 0.90\% & 0.79\%      \\
\hline $\frac{\Delta{\bm\alpha}}{\bm\alpha}=0$, $\frac{\Delta{\bm\beta}}{\bm\beta}=20\%$, $\frac{\Delta{\bf F}}{{\bf F}}=0$ & 2.22\% & 1.19\% & 0.86\%    \\
\hline $\frac{\Delta{\bm\alpha}}{\bm\alpha}=0$, $\frac{\Delta{\bm\beta}}{\bm\beta}=0$, $\frac{\Delta{\bf F}}{{\bf F}}=-20\%$  & 2.41\%  & 1.37\%  & 1.01\%     \\
\hline $\frac{\Delta{\bm\alpha}}{\bm\alpha}=0$, $\frac{\Delta{\bm\beta}}{\bm\beta}=0$, $\frac{\Delta{\bf F}}{{\bf F}}=20\%$ & 1.54\%  & 1.19\%  & 1.03\%      \\
\hline $C_I$ & 3.90\% & 3.37\% & 2.94\%  \\
\hline $C_{II}$ & div & 11.94\% & 7.11\%   \\
\hline
\end{tabular}
\caption{EKF in perfect model scenario - ${\bf P}^m=0$. Time average analysis error variance. Model parametric errors are given in the first column. Assimilation interval is given in the first row. All results are computed over 5 years of simulated time after a 1-year long spin-up period. Error are expressed as percentage of the system climate variance; "div" stands for filter divergence.} 
\end{table}

\begin{table}[hc]
\label{tab3}
\begin{tabular}{||c|c|c|c|c|c||}
\hline\hline
$\frac{\Delta{\bm\lambda}}{{\bm\lambda}}$  & $\tau=12$ hrs & $\tau=6$ hrs & $\tau=3$ hrs  \\ \hline \hline
\hline $\frac{\Delta{\bm\alpha}}{\bm\alpha}=-20\%$, $\frac{\Delta{\bm\beta}}{\bm\beta}=0$, $\frac{\Delta{\bf F}}{{\bf F}}=0$ & 2.97\%  & 2.29\%  & 1.88\%   \\
\hline $\frac{\Delta{\bm\alpha}}{\bm\alpha}=20\%$, $\frac{\Delta{\bm\beta}}{\bm\beta}=0$, $\frac{\Delta{\bf F}}{{\bf F}}=0$ & 2.99\% & 2.51\%  & 2.07\%    \\
\hline  $\frac{\Delta{\bm\alpha}}{\bm\alpha}=0$, $\frac{\Delta{\bm\beta}}{\bm\beta}=-20\%$, $\frac{\Delta{\bf F}}{{\bf F}}=0$ & 1.03\% & 0.86\% & 0.75\%      \\
\hline $\frac{\Delta{\bm\alpha}}{\bm\alpha}=0$, $\frac{\Delta{\bm\beta}}{\bm\beta}=20\%$, $\frac{\Delta{\bf F}}{{\bf F}}=0$ & 1.50\% & 1.05\% & 0.80\%    \\
\hline $\frac{\Delta{\bm\alpha}}{\bm\alpha}=0$, $\frac{\Delta{\bm\beta}}{\bm\beta}=0$, $\frac{\Delta{\bf F}}{{\bf F}}=-20\%$ & 0.92\% & 0.77\% & 0.67\%   \\
\hline $\frac{\Delta{\bm\alpha}}{\bm\alpha}=0$, $\frac{\Delta{\bm\beta}}{\bm\beta}=0$, $\frac{\Delta{\bf F}}{{\bf F}}=20\%$ & 0.91\% & 0.77\% & 0.66\%     \\
\hline $C_I$ & 2.39\% & 1.93\% & 1.69\%  \\
\hline $C_{II}$ & 3.59\% & 3.15\% & 2.75\%   \\
\hline
\end{tabular}
\caption{EKF with white-noise model error assumption - ${\bf P}^m={\bf P}^m_{wn}$. Time mean analysis error variance. The linear, white noise, approximation, Eq. (\ref{PMWNapprox}) is employed. Model parametric errors are given in the first column. Assimilation interval is given in the first row. All results are computed over 5 years of simulated time after a 1-year long spin-up period. Error are expressed as percentage of the system climate variance.} 
\end{table}

\begin{table}[hc]
\label{tab4}
\begin{tabular}{||c|c|c|c|c|c||}
\hline\hline
$\frac{\Delta{\bm\lambda}}{{\bm\lambda}}$  & $\tau=12$ hrs & $\tau=6$ hrs & $\tau=3$ hrs  \\ \hline \hline
\hline $\frac{\Delta{\bm\alpha}}{\bm\alpha}=-20\%$, $\frac{\Delta{\bm\beta}}{\bm\beta}=0$, $\frac{\Delta{\bf F}}{{\bf F}}=0$ & 2.54\%  & 1.89\%  & 1.54\%   \\
\hline $\frac{\Delta{\bm\alpha}}{\bm\alpha}=20\%$, $\frac{\Delta{\bm\beta}}{\bm\beta}=0$, $\frac{\Delta{\bf F}}{{\bf F}}=0$ & 2.66\%  & 2.16\%  & 1.78\%    \\
\hline $\frac{\Delta{\bm\alpha}}{\bm\alpha}=0$, $\frac{\Delta{\bm\beta}}{\bm\beta}=-20\%$, $\frac{\Delta{\bf F}}{{\bf F}}=0$ & 0.97\% & 0.82\%  & 0.72\%      \\
\hline $\frac{\Delta{\bm\alpha}}{\bm\alpha}=0$, $\frac{\Delta{\bm\beta}}{\bm\beta}=20\%$, $\frac{\Delta{\bf F}}{{\bf F}}=0$ & 1.61\% & 1.01\% & 0.77\%    \\
\hline $\frac{\Delta{\bm\alpha}}{\bm\alpha}=0$, $\frac{\Delta{\bm\beta}}{\bm\beta}=0$, $\frac{\Delta{\bf F}}{{\bf F}}=-20\%$ & 0.92\% &  0.77\% &  0.67\%   \\
\hline $\frac{\Delta{\bm\alpha}}{\bm\alpha}=0$, $\frac{\Delta{\bm\beta}}{\bm\beta}=0$, $\frac{\Delta{\bf F}}{{\bf F}}=20\%$ &  0.91\%  &0.77\%  & 0.66\%     \\
\hline $C_I$ & 2.17\% & 1.80\% & 1.60\%  \\
\hline $C_{II}$ & 3.02\% & 2.45\% & 1.99\%   \\
\hline
\end{tabular}
\caption{EKF with deterministic process assumption - ${\bf P}^m={\bf P}^m_{dp}$. Time mean analysis error variance. The quadratic, deterministic noise, approximation, Eq. (\ref{PMapprox}) is employed. Model parametric errors are given in the first column. Assimilation interval is given in the first row. All results are computed over 5 years of simulated time after a 1-year long spin-up period. Error are expressed as percentage of the system climate variance.} 
\end{table}

\newpage

\end{document}